\documentclass[twocolumn]{aastex631}
\usepackage{color}

\usepackage{color}

\newcommand{\HeI}{\makebox{He{\sc I}\,}}

\newcommand{\hbeta}{H{$\beta$}}

\def \OIII {[O\,{\sc iii}]}

\newcommand{\OIIIb}{[O{\sevenrm\,III}]\,$\lambda$5007}
\newcommand{\OIIIab}{[O{\sevenrm\,III}]\,$\lambda\lambda$4959,5007}

   \font\sevenrm=cmr7 scaled 1000

\newcommand{\SIII}{[S{\sevenrm III]}}

\usepackage{soul}

\begin{document}

\title{NEXUS Early Data Release: NIRCam Imaging and WFSS Spectroscopy from the First (Partial) Wide Epoch}

\author[0000-0001-5105-2837]{Ming-Yang Zhuang}
\affiliation{Department of Astronomy, University of Illinois Urbana-Champaign, Urbana, IL 61801, USA}

\author[0000-0002-7633-431X]{Feige Wang}
\affiliation{Department of Astronomy, University of Michigan, 1085 S. University Ave., Ann Arbor, MI 48109, USA}

\author[0000-0002-4622-6617]{Fengwu Sun}
\affiliation{Center for Astrophysics $\vert$ Harvard \& Smithsonian, 60 Garden Street, Cambridge, MA 02138, USA}

\author[0000-0003-1659-7035]{Yue Shen}
\affiliation{Department of Astronomy, University of Illinois Urbana-Champaign, Urbana, IL 61801, USA}
\affiliation{National Center for Supercomputing Applications, University of Illinois Urbana-Champaign, Urbana, IL 61801, USA}

\author[0000-0002-1605-915X]{Junyao Li}
\affiliation{Department of Astronomy, University of Illinois Urbana-Champaign, Urbana, IL 61801, USA}

\author[0000-0002-6523-9536]{Adam J.~Burgasser}
\affiliation{Department of Astronomy \& Astrophysics, UC San Diego, La Jolla, CA, USA}

\author[0000-0003-3310-0131]{Xiaohui Fan} 
\affiliation{Steward Observatory, University of Arizona, Tucson, AZ 85750, USA}

\author[0000-0002-5612-3427]{Jenny E.~Greene}
\affiliation{Department of Astrophysical Sciences, Princeton University, Princeton, NJ 08544, USA}

\author[0000-0001-6022-0484]{Gautham Narayan}
\affiliation{Department of Astronomy, University of Illinois Urbana-Champaign, Urbana, IL 61801, USA}
\affiliation{Center for AstroPhysical Surveys, National Center for Supercomputing Applications, University of Illinois Urbana-Champaign, Urbana, IL 61801, USA}

\author[0000-0003-3509-4855]{Alice E.~Shapley}
\affiliation{Department of Physics \& Astronomy, University of California, Los Angeles, 430 Portola Plaza, Los Angeles, CA 90095, USA}

\author[0000-0002-6893-3742]{Qian Yang}
\affiliation{Center for Astrophysics $\vert$ Harvard \& Smithsonian, 60 Garden Street, Cambridge, MA 02138, USA}

\begin{abstract}

We present the Early Data Release of the Multi-Cycle JWST-NEXUS Treasury program (2024-2028), which includes NIRCam imaging and WFSS observations from the first (partial) NEXUS-Wide epoch covering the central 100~${\rm arcmin^2}$ of the NEXUS field, located near the North Ecliptic Pole and within the Euclid Ultra-Deep Field. We release reduced NIRCam mosaics (F090W, F115W, F150W, F200W, F356W, F444W), photometric source catalogs, as well as preliminary WFSS spectra (in F322W2 and F444W) for the subset of bright sources (F356W$<$21 mag or F444W$<$21 mag). These observations fully cover the NEXUS-Deep area, and anchor the long-term baseline of the program. These data will be used for initial target selection for the NIRSpec/MSA spectroscopy starting from June 2025. The NIRCam imaging reaches depths of 27.4--28.2 (AB) mags in F090W--F444W. Upcoming NEXUS-Wide epochs will expand the area to the full $\sim 400\,{\rm arcmin^2}$, and improve the NIRCam exposure depths in the Wide tier by a factor of three. In addition, this central region will be repeatedly covered by the NEXUS-Deep observations (NIRCam imaging and NIRSpec/MSA PRISM spectroscopy) over 18 epochs with a $\sim 2$-month cadence. We demonstrate the data quality of the first NEXUS observations, and showcase some example science cases enabled by these data. 
\end{abstract}


\section{Introduction} \label{sec:intro}
The James Webb Space Telescope \cite[JWST;][]{Gardner2006SSRv} has enabled unprecedented observations of the distant and faint Universe. Medium to large survey programs, such as JADES \citep{JADES}, UNCOVER \citep{UNCOVER}, COSMOS-Web \citep{COSMOS-Web}, FRESCO \citep{FRESCO}, PRIMER \citep{PRIMER}, NGDEEP \citep{NGDEEP}, EIGER \citep{EIGER}, ASPIRE \citep{Wang23}, CEERS \citep{CEERS}, PEARLS \citep{PEARLS}, GLASS-JWST \citep{GLASS}, etc., from the first few cycles have already produced a wealth of IR imaging and spectroscopic data.
These surveys have been addressing many important questions in extragalactic astronomy, such as the formation of the first galaxies and supermassive black holes \citep[e.g.,][]{2023NatAs...7..622C, 2024Natur.627...59M, 2024Natur.633..318C, 2024arXiv240920549S} and the start of the quenching of massive galaxies at $z>4$ \citep[e.g.,][]{2023Natur.619..716C, 2024MNRAS.534..325C, 2024arXiv240405683D}.

The North ecliptic pole EXtragalactic Unified Survey (NEXUS) project \citep{nexus} is a multi-cycle (3, 4, and 5) Treasury program (368 hrs total) that monitors a field in the JWST Continuous Viewing Zone (CVZ). The footprint of the NEXUS survey is chosen to be within the Euclid Ultra-Deep Field \citep{euclid}. NEXUS has two tiers of area coverage and cadences. The Wide tier (400 arcmin$^2$) has three epochs, each of which costs $\sim$65 hrs, to be observed with primary NIRCam/WFSS at three PAs (PA1, PA1+135 deg, PA1+270 deg). The Deep tier ($\sim$50 arcmin$^2$ within Wide) has 18 epochs, each with $\sim$10 hrs, performing primary NIRCam imaging and primary NIRSpec MOS spectroscopy with a cadence of $\sim$2 months over three years (2024--2028). In each tier, NIRCam and MIRI imaging is also obtained either as primary or coordinated parallel observations. The full details about the design of the NEXUS program, the scheduling of the observations, and science motivation are presented in the overview paper \citep{nexus}. 

In this paper, we present the early data release (EDR) of NIRCam imaging and WFSS from NEXUS-Wide tier epoch1 observations, covering the central 100 arcmin$^2$. This release includes fully reduced NIRCam images in the F090W, F115W, F150W, F200W, F356W, and F444W filters, a photometric source catalog, and reduced NIRCam WFSS 1D spectra of bright sources (F356W$<$21 mag or F444W$<$21 mag) in F322W2 and F444W grisms. The paper is structured as follows. We describe the NEXUS observations in Section~\ref{sec2}, introduce the data reduction methodology in Section~\ref{sec3}, photometric catalog construction in Section~\ref{sec4}, basic data properties in Section~\ref{sec5}, and data products and access in Section~\ref{sec6}. 

\begin{figure*}[t]
\centering
\includegraphics[width=0.9\textwidth]{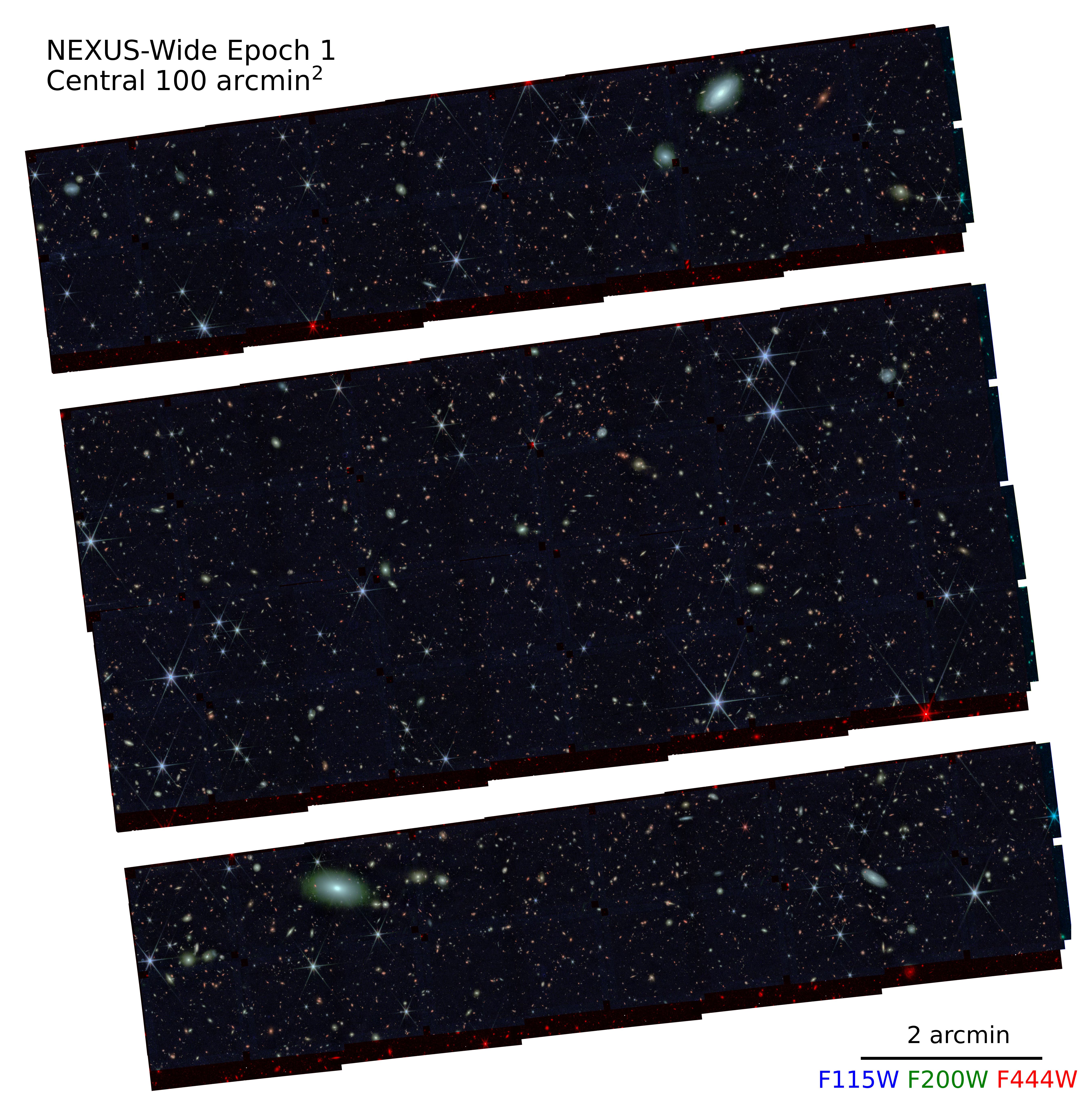}
\caption{Pseudo-color image of the NEXUS-Wide Epoch1 NIRCam observations (covering the central 100 arcmin$^2$), constructed from the F444W (red), F200W (green), and F115W (blue) images. The scale bar indicates two arcmin. North is up and east is left. The image gaps in the first Wide epoch will be covered by future epochs. {Certain small patches and edges of the mosaics are only covered by F444W in the first epoch and are shown in monochromatic red color.}}
\label{fig1}
\end{figure*}

\section{NEXUS Implementation and First Observations}\label{sec2}

NEXUS-Wide observations perform NIRCam/WFSS \citep{Rieke23} as primary and MIRI/imaging \citep{Wright23} as parallel. Each epoch observation uses both NIRCam Modules and adopts a $4\times9$ (column $\times$ row) mosaic pattern, covering a total area of $\sim$360 arcmin$^2$. Each epoch of NEXUS-Deep tier observations combines a $2\times3$ NIRCam/imaging mosaic covering the central $\sim$70 arcmin$^2$ paired with MIRI/imaging, and a $2\times2$ NIRSpec/MOS mosaic covering the central $\sim$49 arcmin$^2$ paired with NIRCam/imaging parallel.

NEXUS-Wide tier observations adopt a combination of F322W2 and F444W filters in the long wavelength (LW) channel of NIRCam/WFSS observations, paired with F115W and F200W imaging in the short wavelength (SW) channel to cover a complete wavelength range from $\sim$2.4 to 5 micron with grism spectroscopy. F150W+F356W (in-field) and F090W+F444W (both in-field and out-field) direct imaging is obtained after WFSS exposures. We only use Grism-R for WFSS observations since the source confusion will be resolved by the detection of multiple emission lines and three different PAs provided by three epochs. We adopt a two-point \texttt{INTRAMODULEX} primary dither pattern to fill the gaps between SW detectors and mitigate the effect of cosmic rays. To reduce overhead, we do not require subpixel dither, as we will have three epoch observations with different PAs. We adopt the \texttt{SHALLOW4} readout pattern with 6 groups and 1 integration for F322W2 WFSS exposures and the \texttt{MEDIUM8} readout pattern with 6 groups and 1 integration for F444W WFSS exposures. All direct imaging exposures are obtained with the \texttt{SHALLOW4} readout pattern with 6 groups and 1 integration. 

The above configurations will result in a total exposure time of 623s for F322W2 WFSS observations, 1245s for F444W WFSS observations, 311s for F150W+F356W direct imaging, and 934s for F090W+F444W direct imaging (in-field + out-field) per epoch. For NEXUS-Wide tier epochs 2 and 3, we only require one set of direct imaging (in-field + out-field), with F150W+356W for epoch 2 and F090W+F444W for epoch 3. {This setting ensures we will have more uniform imaging exposure depths across NIRCam filters in the final data, but it results in non-uniform depths in earlier epochs.} For epoch 1, we only have 1 exposure for F150W and F356W imaging without any dither, which would leave empty pixels in the mosaic. As described in Section~\ref{sec4.2}, source photometry is measured in point spread function (PSF)-homogenized images after convolution. By adopting \texttt{astropy}'s convolution, which is able to replace NaN based on interpolation, we will recover most of the flux since cosmic rays/bad pixels only affect 0.4\% and 1.9\% of the total pixels in the F150W and F356W filters, respectively. This issue will be completely resolved by combining data from NEXUS-Wide epoch 2 observations, which have 3 dithers. 

Due to the requirement of JWST's Long Range Plan, Wide tier epoch 1 observations are divided into two scheduled blocks. The first block of observations were executed during Sep 12--16 2024, consisting of the central $2\times5$ tiles with a coverage of $\sim$100 arcmin$^2$ (totaling $\sim 18$\,hrs). The second part of the observations will be executed in mid June 2025 to complete the remaining 26 tiles. NEXUS-Deep tier observations will start from early June 2025 and repeat with a cadence of $\sim$2 months and a PA rotation of $\sim$60 deg. The split of the Wide epoch 1 observations has the following benefits: (1) provide the reference image and catalog for transients and NIRSpec/MOS target selection; (2) effectively extend our fiducial time baseline in the NEXUS-Deep area by nine months to a total of 43 months; (3) allow a comprehensive assessment of the performance of F322W2 filter in extragalactic NIRCam/WFSS programs; (4) have adequate time ahead of the bulk of the Deep observations to enable community coordination on follow-up multi-wavelength observations. 

\section{Data Reduction} \label{sec3}

In this EDR, we reduce all NIRCam data from the partial Wide tier epoch 1 observations executed in Sep 2024. This dataset contains $2\times5$ tiles covering the central $\sim$100 arcmin$^2$ with NIRCam/imaging in the F090W, F115W, F150W, F200W, F356W, and F444W filters and NIRCam/WFSS with the F322W2 and F444W grisms. {Parallel MIRI imaging data will be presented in the NEXUS Data Release 1 (DR1), scheduled in late 2025/early 2026 \citep{nexus}. }

\subsection{NIRCam Imaging}
The data reduction procedures largely follow those presented in \citet{2024ApJ...962...93Z} and are similar to other JWST programs (e.g., CEERS and JADES). We reduce the data using the latest \texttt{jwst}\footnote{\url{https://jwst-pipeline.readthedocs.io/en/latest/}} calibration pipeline \citep[version 1.15.1;][]{Bushouse_JWST_Calibration_Pipeline_2024} and the latest Calibration Reference Data System (CRDS) with context file \texttt{jwst\_1281.pmap} with custom steps and modifications. The overview of our reduction procedures is presented below.

\subsubsection{Stage 1 - Detector-level Corrections}
The uncalibrated raw images of individual exposures are retrieved from MAST\footnote{\url{https://mast.stsci.edu/}} and fed to stage 1 pipeline \texttt{Detector1Pipeline}. This step applies basic detection-level corrections to all exposure types and performs ramp fitting afterwards. We adopt the default configuration except setting \texttt{expand\_large\_events=True} in the \texttt{Jump Detection} step to turning on large cosmic ray events (snowballs) identification and flagging. 

\subsubsection{Stage 2 - Imaging Calibration}

NIRCam/imaging exposures are fed to stage 2 pipeline \texttt{Image2Pipeline}. 
This step includes wcs assignment, flat-fielding, and photometric calibration and returns fully calibrated individual exposures. Stage 2 pipeline is run with default parameters. 

We then remove the large-scale background and $1/f$ noise (horizontal and vertical patterns) in individual exposures. The source mask is generated in smoothed images (with a Gaussian kernel with full-width-at-half-maximum of 3 pixels) with a threshold of 2$\sigma$ and a dilation radius of 10 pixels for short-wavelength (SW) filters and 5 pixels for long-wavelength (LW) filters. After initial source mask is generated, we measure 2-dimensional background in source emission-masked images with a box size of $256\times 256$ pixels and a filter size of $3\times3$ boxes using \texttt{SExtractorBackground} implemented in \texttt{photutils}. We repeat the above source masking and background estimate steps two times to obtain the final source mask and background model. $1/f$ noise is measured in background-subtracted, source-masked images in an amplifier-by-amplifier manner. 

Our NIRCam exposures are affected by various scattered light artifacts, including ``wisps'', large area parallel striping, and type II Dragon's breath features\footnote{\url{https://jwst-docs.stsci.edu/known-issues-with-jwst-data/nircam-known-issues/nircam-scattered-light-artifacts}}. ``wisp'' features are caused by scattered light coming off-axis and bouncing off the top secondary mirror strut. They are stationary features present in the same detector locations in all exposures with variable brightness in different filters. They are most prominent in the B4 detector, with fainter features in A3, A4, and B3 detectors. Wisp templates are created by stacking 2$\sigma$-clipped, source emission masked-images in each detector and each band. We directly subtract the corresponding wisp template from each image without scaling its amplitude. Large area parallel striping features are observed in the first visit of all SW filters (most prominent in the F115W filter), which are expected to be caused by scattered light or a diffraction spike from a very bright source located far from the detectors' field of view. These large area parallel stripes extend across the b3 detector and affect the upper-right corner of b4 detector (together with faint diffuse emission). We construct a template for each detector by median stacking data/error images from all the affected SW exposures after masking bright sources and a 2$\sigma$-clipping. A best-fit scaling factor is multiplied to the template before subtracting the striping features in individual exposures. Type II dragon's breath is caused by light scattering off a knife edge installed to block a stray light path to the short wavelength detectors. It affects only one exposure of F115W, F150W, and F200W filters. We simply mask the affected pixels.

We find that \texttt{outlier\_detection} in Stage 3 pipeline fails to reject some hot/warm/cosmic ray-affect pixels in mosaics with few dithers. Therefore, we apply additional outlier rejection prior to mosaic construction step. Bad pixels that have persist detector positions are removed by subtracting the median source-masked images in each detector and each band. We do not repeat the subtraction for detectors and filters involved in the wisp removal step as mentioned above. We select other isolated bad pixels that have high SNR ($>15$) together with low SNR ($<5$) or negative SNR ($<-10$) connected pixels. Finally, we mask the remaining cosmic-ray affected pixels that are not properly identified and flagged by Stage 1 \texttt{Jump Detection} step. These are mainly clusters of connected pixels with large fluxes and errors. We first select sources with at least five connected pixels above 5$\sigma$. Cosmic-ray affected pixels are well separated with real sources by having surface brightness within the detection segment (flux/area) much larger than 50 (10) in SW (LW) filters. We mask these pixels by dilating their segments with a radius of 3 pixels. A small amount of pixels that are high above the sky standard deviation ($>5\sigma$) but low compared to their errors ($<2\sigma$) are also masked. 

\subsubsection{Stage 3 - Mosaic Construction}
we run Stage 3 pipeline \texttt{Image3Pipeline} to produce a single mosaic for each filter by combining all the calibrated images (all dithers and detectors) from all ten tiles observed so far. We turn off \texttt{Skymatch} step since we have subtracted the sky in the previous step. We adopt \texttt{tweakreg} step for astrometry correction, \texttt{outlier\_detection} step for additional outlier pixel rejection, and \texttt{resample} step for image resample and combination. Astrometry is tied to the Subaru Hyper Suprime-Cam (HSC) catalog of The Hawaii eROSITA Ecliptic Pole Survey \citep[HEROES;][]{2023ApJS..266...24T}, which was calibrated to Pan-STARRS. Instead of grouping exposures from the same visit together, we group exposures from the same Module to account for uncertainties in the locations of the detectors relative to one another in the focal plane\footnote{\url{https://jwst-docs.stsci.edu/known-issues-with-jwst-data/nircam-known-issues/nircam-imaging-known-issues}}. We adopt the default configurations for \texttt{outlier\_detection} except for setting \texttt{maskpt=0.5}, which is the fraction of maximum weight to use as the lower limit for valid data, to account for weight variance within and across detectors, as suggested in \citet{2024ApJ...962...93Z}. For the \texttt{resample} step, we adopt a pixel scale (\texttt{pixel\_scale}) of 0\farcs03 pixel$^{-1}$, and input pixel ``shrunk'' fraction (\texttt{pixfrac}) of 0.8 for all the filters. All the other parameters are kept as their default values. All the images are registered to the same reference coordinates to enable convenient pixel-level analysis. A final round of background subtraction is performed to the mosaic using the same configuration and procedure described above to remove the residual large scale background. We find a small residual offset ($\Delta{\rm RA}=-0\farcs0127$ and $\Delta{\rm DEC}=0\farcs0015$) and directly correct it in the final mosaic. 

We rescale the weight and error maps using the ratios between the $\sigma$ from gaussian fitting to the background dominated pixels in the data and that in the weight and error maps, respectively. We derive scaling factors for areas covered by different numbers of exposure separately. The scaling factors vary from 0.63 to 1.46 for inverse root of weight maps and from 0.61 to 0.97 for error maps.

\begin{figure}[t]
\centering
\includegraphics[width=0.5\textwidth]{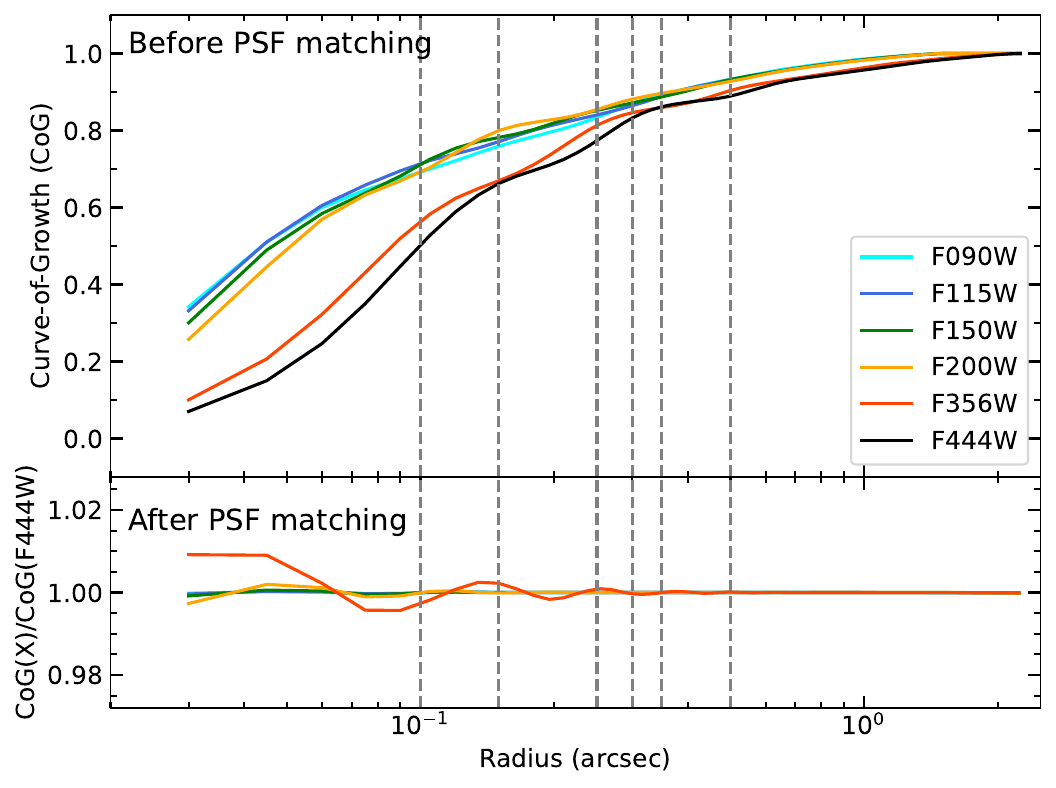}
\caption{Top: Normalized curve-of-growth (CoG) of the encircled light over total tight as a function of radius for NIRCam PSF models in the F090W (cyan), F115W (blue), F150W (green), F200W (oragne), F356W (red), and F444W (black) filters. Bottom: The ratios of CoG in the other filters and that in the F444W filter after PSF matching (see Section~\ref{sec4.1}). Gray vertical dashed lines indicate the sizes of our fixed aperture photometry.}
\label{fig2}
\end{figure}

\subsection{NIRCam WFSS Data Reduction}\label{wfss_reduction}

The NIRCam/WFSS data were reduced using the combination of \texttt{jwst.calwebb\_detector1} (version 1.10.2; \citealt{Bushouse2023}) and some custom scripts. The calibration reference files (\texttt{jwst\_1281.pmap}) were obtained from the standard Calibration Reference Data System (CRDS, version 11.17.19). The detailed description of the custom scripts can be found in \citet{Wang23} and \citet{Yang23}. Briefly, we used the standard \texttt{calwebb\_detector1} stage 1 pipeline to produce calibrated rate files. We then subtracted $1/f$ noise pattern only along columns because the spectra are dispersed along detector rows. To correct the astrometric offsets between individual WFSS exposures and the mosaic images, we determined the offsets between each of the SW images obtained simultaneously with WFSS exposures and the mosaic image. The measured astrometric offsets were then applied to the tracing and dispersion models\footnote{\href{https://github.com/fengwusun/nircam_grism}{https://github.com/fengwusun/nircam\_grism}} \citep{Sun23} in later stages when extracting the spectra. 
We note the tracing and dispersion models have been updated using calibration data taken through commissioning, Cycle-1 and 2 (up to June 2024), and the accuracy has been further improved with negligible wavelength calibration error ($\Delta v \lesssim 40$ km s$^{-1}$) and tracing offset ($\lesssim 0\farcs01$) for most commonly used filters including F322W2 and F444W (Sun, F.\ et al.\ in prep.).
The WFSS sky background is wavelength dependent. In order to subtract the background, we constructed master background models for F444W and F322W2 observations separately using all NEXUS exposures. 

We extract both 2D and 1D spectra of all sources detected in the mosaic image (see Section \ref{sec4} for more details about the source catalog construction). The 2D spectrum of each source is extracted from individual exposure and then coadded into a single 2D spectrum following the histogram2D technique. We then extracted 1D spectra from the coadded 2D spectra using both optimal \citep{Horne86} and boxcar extraction algorithms. Additionally, we optimally extracted 1D spectra from all individual 2D spectra. The extracted 1D spectra from individual exposures were then combined to a stacked 1D spectrum for each source. The main purpose for producing a second version of stacked 1D spectra is that such method provides a better removal of outliers for bright continuum sources. 
All grism spectra are flux-calibrated using grism sensitivity curves constructed through Cycle-1 flux calibration observations (PID: 1536, 1537, 1538; see \citealt{Sun23}), and the flux calibration uncertainty is $2-3\%$.

\section{Catalog Construction} \label{sec4}
\subsection{Detection}\label{sec4.1}
We create noise-equalized mosaic by coadding the square-root of the inverse variance-weighted F356W and F444W images. This coadded image is used as the deep detection image, maximizing the detectability of faint high-redshift sources. Since the variable noise across the mosaic has been taken into account during detection image construction, we use a weight map filled with ones and zeros, with one and zero indicating the area with and without coverage, respectively. We fill the saturated core of saturated stars and smooth central high surface-brightness pixels to reduce the detection of spurious sources and to help the deblending of close faint sources. We use \texttt{Source Extractor} \citep[\texttt{SExtractor};][]{1996A&AS..117..393B} double-image mode to perform source detection on detection image and analysis on individual PSF-homogenized images (see Section~\ref{sec4.2}). We use a relative detection and analysis threshold of 1.5 $\sigma$ with a minimum area of 15 pixels. A $7\times7$ Gaussian filter with a FWHM=4 pixels is used to smooth the images before detection. We adopt a deblending threshold of 32 with a minimum contrast parameter of 0.005 and turn on spurious sources cleaning with a \texttt{CLEAN\_PARAM=1}.

\begin{deluxetable}{cccc}[ht]
\caption{NEXUS-Wide Epoch1 Central Mosaic Properties \label{table1}}
\tabletypesize{\small}
\tablehead{
\colhead{Filter} & \colhead{Exposure Time} & \colhead{Area} & \colhead{5$\sigma$ Depth} \\
\colhead{} & \colhead{[s]} & \colhead{[arcmin$^2$]} & \colhead{[mag ($n$Jy)]}
}
\startdata
F090W & 311.4 & 13.5 & 26.67 (77.6) \\ 
F090W & 622.7 & 24.1 & 27.19 (48.5) \\
F090W & 934.1 & 58.3 & 27.38 (40.6) \\
F090W & 1245.5 & 1.9 & 27.43 (38.6) \\
F090W & 1556.8 & 1.5 & 27.67 (31.0) \\
F090W & 1868.2 & 2.7 & 27.91 (24.8) \\
\hline
F115W & 311.4 & 20.4 & 26.81 (68.3) \\ 
F115W & 622.7 & 66.5 & 27.20 (47.8) \\
F115W & 934.1 & 6.2 & 27.46 (37.7) \\
F115W & 1245.5 & 1.7 & 27.64 (31.9) \\
\hline
F150W & 311.4 & 84.2 & 27.15 (50.2) \\ 
\hline
F200W & 622.7 & 20.4 & 27.76 (28.5) \\ 
F200W & 1245.5 & 66.4 & 28.08 (21.4) \\ 
F200W & 1868.2 & 6.4 & 28.23 (18.5) \\ 
F200W & 2490.9 & 1.7 & 28.53 (14.1) \\ 
\hline
F356W & 311.4 & 87.3 & 27.91 (24.9) \\ 
\hline
F444W & 311.4 & 12.4 & 27.50 (36.3) \\ 
F444W & 622.7 & 17.8 & 27.77 (28.4) \\
F444W & 934.1 & 67.2 & 28.22 (18.7) \\
F444W & 1245.5 & 1.4 & 28.38 (16.2) \\
F444W & 1556.8 & 1.1 & 28.32 (17.1) \\
F444W & 1868.2 & 2.1 & 28.54 (13.9) \\
\enddata
\tablecomments{Basic properties of the NEXUS-Wide Epoch1 central mosaics. Regions with longer than nominal exposure times (Section~\ref{sec5.1}) are due to overlaps from dithered exposures. 5$\sigma$ depth is measured from randomly placed 0\farcs3-diameter apertures without aperture correction. The fractional fluxes within an 0\farcs3-diameter aperture are 0.744, 0.769, 0.783, 0.796, 0.621, and 0.675 compared to the total fluxes of point sources for F090W, F115W, F150W, F200W, F356W, and F444W filters, respectively.}
\end{deluxetable}

\begin{figure*}[t]
\centering
\includegraphics[width=0.9\textwidth]{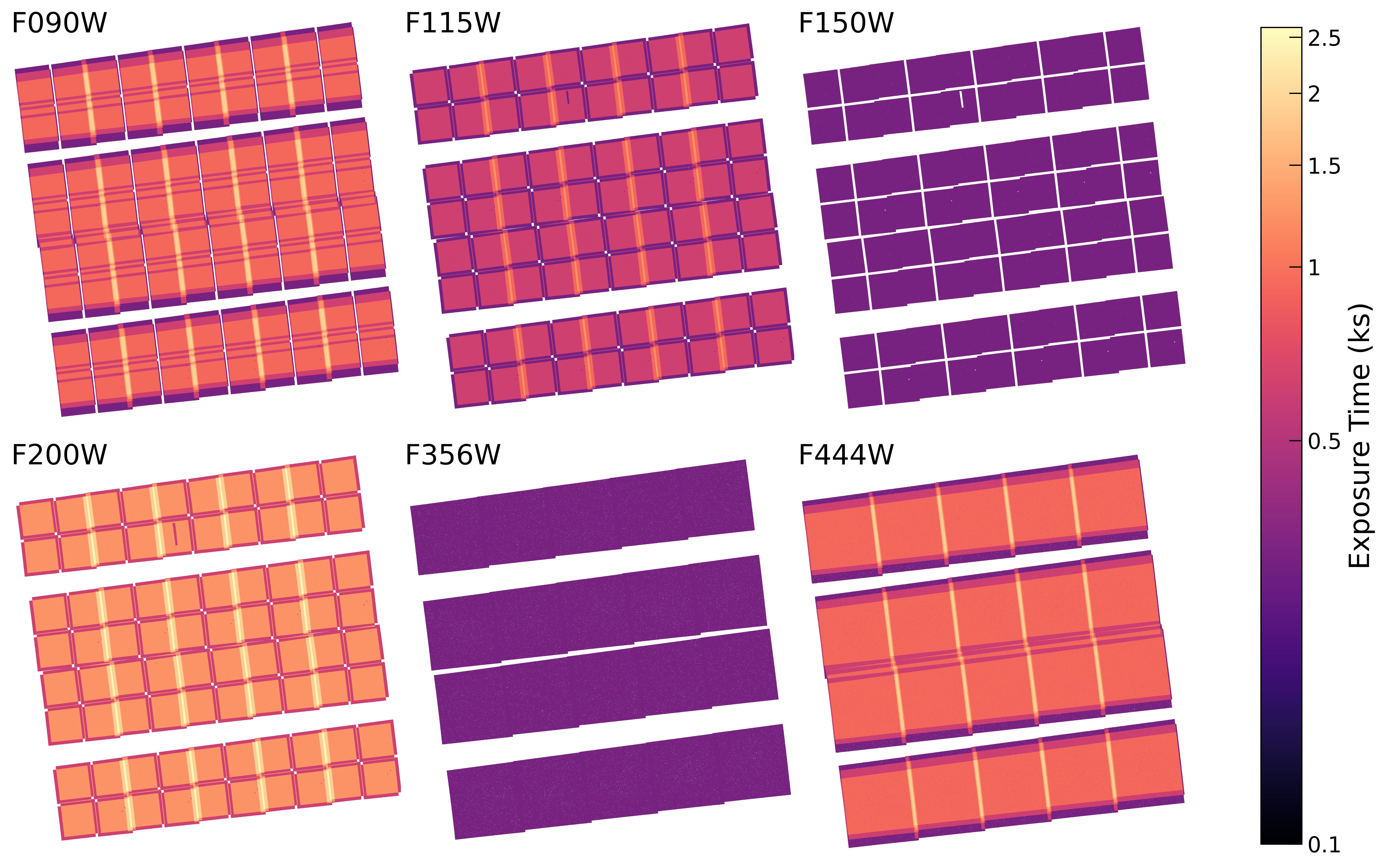}
\caption{Exposure time maps of the NEXUS-Wide Epoch1 central mosaics, with brighter colors indicating more exposure time as shown in the color bar. 
Note that while gaps exist in the Epoch1 mosaics, most of them will be filled by exposures from NEXUS-Wide epochs 2-3 and NEXUS-Deep epochs.}
\label{fig_expt}
\end{figure*}

\subsection{Photometry}\label{sec4.2}
We construct PSF models using \texttt{PSFEx} \citep{PSFEx} following the procedures described in \citet{2024ApJ...962..139Z, 2024ApJ...962...93Z}. We adopt the \textit{global} PSF model built from all bright, isolated, unsaturated stars in our field-of-view (FoV) as the representative PSF in each filter. We match the PSF of all images to that in the F444W filter with the PSF matching kernels generated using \texttt{pypher} \citep{pypher}. The matched PSF models from the other filters generally agree well with that in the F444W filter, with $\lesssim1$\% deviations in encircled energy curves at $r\leq 2\farcs25$ (Figure~\ref{fig2}). We provide circular aperture fluxes with radius = 0\farcs1, 0\farcs15, 0\farcs25, 0\farcs3, 0\farcs35, and 0\farcs5, and AUTO fluxes for each filter. Common aperture fluxes in PSF-matched images are useful for color measurement and spectral energy distribution fitting. \texttt{SExtractor} AUTO flux is measured within an adaptively scaled elliptical aperture based on the Kron radius \citep{Kron1980}, which typically encloses 90\%--95\% of the total light. We adopt the default \texttt{PHOT\_AUTOPARAMS} with a Kron parameter of 2.5 and a minimum radius of 2.0. Aperture corrections of fluxes outside Kron aperture is measured using the F444W PSF model by calculating the ratio between the encircled flux within the aperture and the total flux within a diameter of 151 pixels ($\sim$4\farcs5) for each source. The reported AUTO flux has been corrected for aperture loss. 

It has been widely acknowledged that the flux errors returned by \texttt{SExtractor} in drizzled images are underestimated due to pixel correlation \citep{2000AJ....120.2747C}. We update the errors of aperture fluxes by measuring $\sigma$ from fitting a Gaussian function to the histogram of fluxes within empty, randomly placed apertures following previous works \citep[e.g.,][]{2005ApJ...624L..81L, 2011ApJ...735...86W, 2014ApJS..214...24S, 2023ApJS..269...16R}. We measure a range of apertures from r=0\farcs1 to 1\farcs2, with 1 aperture per 5000 pixels ($\sim$4.5 arcsec$^2$; at least 2000 apertures) for each number of exposures. We fit a power-law function to the ratio of $\sigma$ in apertures and that of individual pixels and the aperture diameter in each filter of different numbers of exposures separately. The best-fit power-law index is $\sim1.4$ for our mosaics, which is similar to that in JADES mosaics \citep{2023ApJS..269...16R}. The error of AUTO flux is estimated using the best-fit power-law function with circularized Kron diameter. The total error of aperture photometry consists of aperture corrected Poisson uncertainty associated with the source (corrected for aperture loss for AUTO flux) and this updated aperture-based background uncertainty.

Since the Kron aperture is determined from the F356W+F444W detection image, it may include many background dominated pixels in filters where the source is faint and lead to very large AUTO flux uncertainty. Therefore, for the purpose of photometric redshift estimation and spectral energy distribution modeling, it is often useful to define a ``total'' flux (\texttt{FLUX\_TOTAL}) in each filter based on the aperture flux (\texttt{FLUX\_APER}) and a scaling factor ($f_{\rm total}$), assuming no color gradient outside the aperture. $f_{\rm total}$ is derived from the ratio between \texttt{FLUX\_AUTO} and \texttt{FLUX\_APER} in the F444W filter: $f_{\rm total}\equiv$ \texttt{FLUX\_AUTO\_F444W}/\texttt{FLUX\_APER\_F444W}. Therefore, \texttt{FLUX\_TOTAL\_BAND} $=f_{\rm total} \times$ \texttt{FLUX\_APER\_BAND}, with its corresponding error \texttt{FLUXERR\_TOTAL} $=f_{\rm total} \times$ \texttt{FLUXERR\_APER}. For each filter, we derive six \texttt{FLUX\_TOTAL}, one for each aperture.

\begin{figure}[t]
\centering
\includegraphics[width=0.5\textwidth]{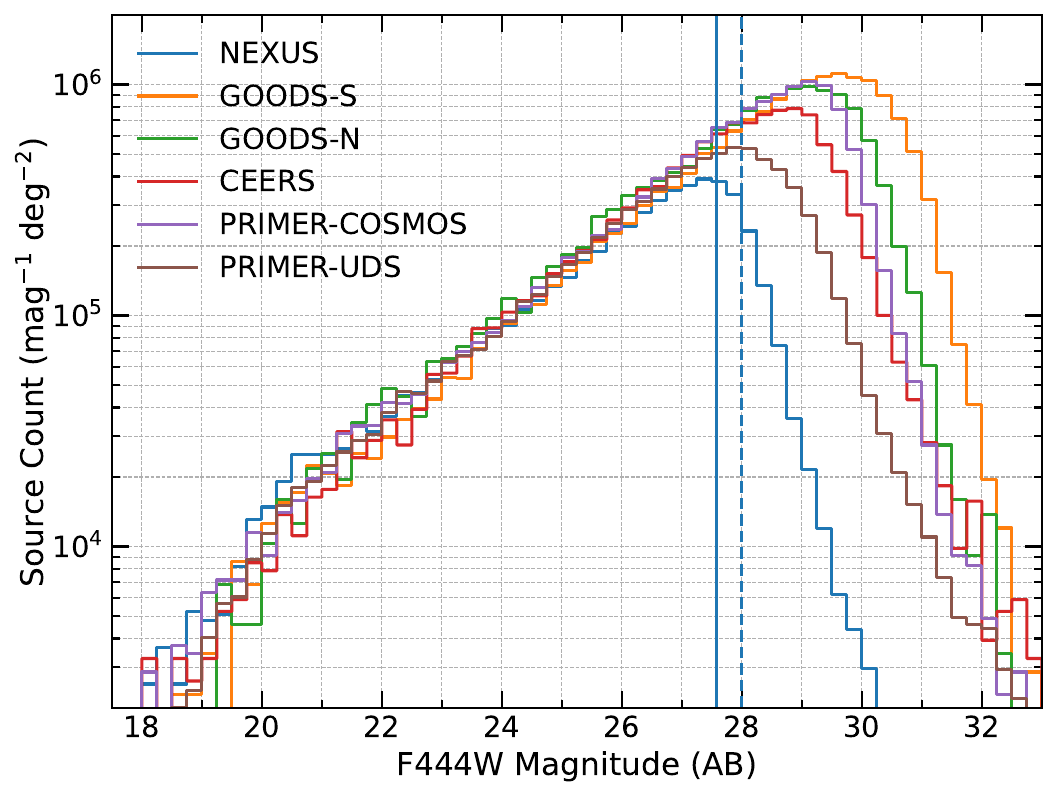}
\caption{Source surface density versus F444W magnitude for objects from NEXUS (blue), GOODS-S (orange), GOODS-N (green), CEERS (red), PRIMER-COSMOS (purple), and PRIMER-UDS (brown). Source counts of other fields are from \cite{2024arXiv240900169M}. Solid and dashed vertical lines indicate area weighted, 5$\sigma$ NEXUS limiting depths with and without aperture correction, respectively.}
\label{fig_mag_hist}
\end{figure}

\section{Early Release Data} \label{sec5}
\subsection{Exposure time, coverage, and depth} \label{sec5.1}
Figure~\ref{fig_expt} shows the exposure time maps of our NIRCam mosaics in the six filters. The exposure times of the majority of our mosaics vary from 311s to 1245s, consisting of 1 to 3 exposures (dithers). F150W and F356W have only one exposure of 311s in NEXUS-Wide epoch 1 observations. Improved mosaics in these two filters with better outlier rejection and spatial sampling will be made after Wide epoch 2 observations (three dithers). The gap between detectors for SW filters and between modules for SW+LW filters will be largely filled by Wide epoch 2 and Deep observations. Table~\ref{table1} shows the basic properties of the mosaics. F090W and F444W have the largest sky coverage ($\sim$102 arcmin$^2$). F115W and F200W have $\sim$95 arcmin$^2$, while F150W and F356W have $\sim$84 arcmin$^2$ and $\sim$87 arcmin$^2$, respectively. We note that a small fraction (5\%--8\%) of the mosaics have up to twice the nominal exposure time and hence better depths.

\subsection{Source counts}
Figure~\ref{fig_mag_hist} shows the number of objects as a function of F444W AUTO magnitude in the NEXUS field. We find a similar distribution as other fields with much deeper NIRCam imaging, including GOODS-S, GOODS-N, CEERS, PRIMER-COSMOS, and PRIMER-UDS \citep{2024arXiv240900169M}, with small deviations owing to cosmic variance. 

We estimate the completeness of our catalog by fitting a power law function to the source counts between 22 and 27 mag. We find 84\%, 50\%, and 16\% completeness of 27.5, 28.0, and 28.5 mag, respectively. As shown in \cite{2013ApJS..207...24G}, completeness estimated from power-law models is consistent with that measured from detecting simulated sources in real images. Our power law-based completeness also agrees with that estimated by dividing our source counts by those in GOODS-S field (27.3, 27.9, and 28.4 mag). 50\% completeness of our catalog is close to the area weighted 5$\sigma$ limiting depth without aperture correction. 

\begin{figure*}[!th]
\centering
\includegraphics[width=\linewidth]{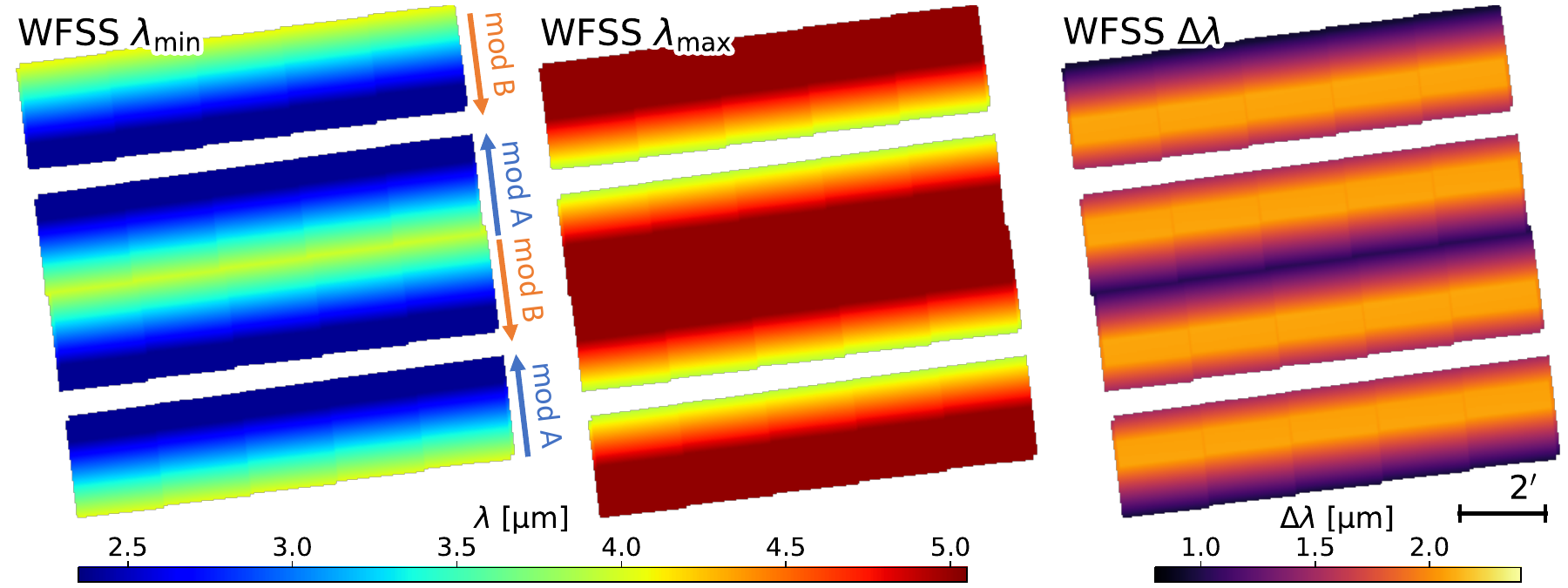}
\caption{NIRCam WFSS wavelength coverage map (minimum $\lambda_\mathrm{min}$, maximum $\lambda_\mathrm{max}$ and  $\Delta \lambda = \lambda_\mathrm{max} - \lambda_\mathrm{min}$) of NEXUS-Wide epoch 1 central mosaics by combining F322W2 and F444W grism data. WFSS dispersion directions on NIRCam module A (blue) and B (orange) are indicated by the arrows.
}
\label{fig:wfss-wave}
\end{figure*}

\begin{figure}[!ht]
\centering
\includegraphics[width=\linewidth]{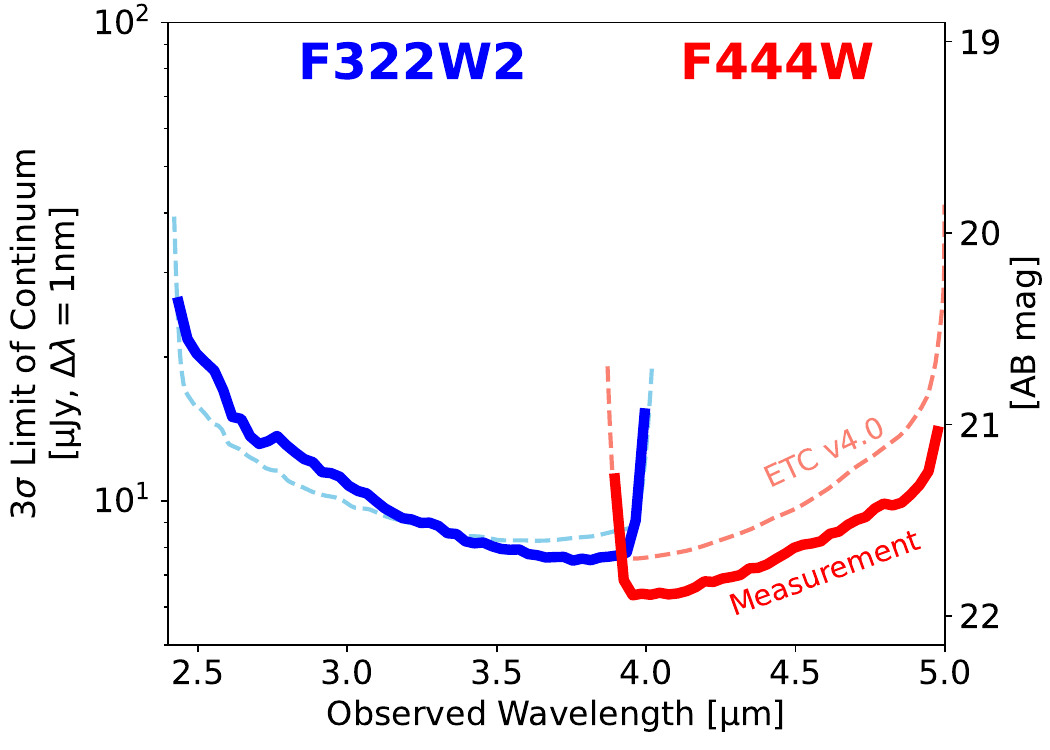}
\caption{NIRCam WFSS continuum sensitivity ($3\sigma$, per $\Delta\lambda = 1$\,nm\,$\sim$\,1\,native pixel) as a function of wavelength for NEXUS F322W2 and F444W Module A Grism-R spectroscopy. 
Solid lines indicate sensitivities that we measured from the NEXUS observations of compact sources at characteristic depths (623\,s for F322W2 and 1245\,s for F444W).
Dashed lines indicate sensitivities from JWST ETC v4.0 assuming conditions similar to the actual observations.
}
\label{fig:wfss-sens}
\end{figure}

\subsection{NIRCam WFSS wavelength coverage}

The effective FoV of NIRCam WFSS is a function of wavelength. 
At the so-called zero deflection wavelength ($\lambda_0 =$\,3.94\,\micron; \citealt{Rieke23}), the effective FoV of WFSS almost equals to the FoV of imaging.
At a larger wavelength offset from $\lambda_0$, spectral pixels ($x_s, y_s$ on detector) are dispersed further away from the direct imaging position of sources ($x_0, y_0$ on detector), making the effective FoV of WFSS smaller.

Figure~\ref{fig:wfss-wave} shows the wavelength coverage of NEXUS-Wide epoch 1 central mosaics, including the minimal wavelength ($\lambda_\mathrm{min}$ down to 2.4\,\micron), maximal wavelength ($\lambda_\mathrm{max}$ up to 5.0\,\micron) and spectral length ($\Delta \lambda = \lambda_\mathrm{max} - \lambda_\mathrm{min}$).
Because we obtained both F322W2 (2.4--4.0\,\micron) and F444W (3.9--5.0\,\micron) grism with the same NIRCam pointing, the spectra of the same sources in two bands can be concatenated to cover a longer wavelength range.
The dataset covers a total area of 100\,arcmin$^2$ with minimal--median--maximal spectral length of 1.0--1.8--2.1\,\micron, respectively.
The spectral length of NEXUS is larger than most of existing NIRCam WFSS surveys which primarily used one wide-band filters.
A long spectral length will be useful to obtain $\geq2$ lines detections for secure spectroscopic redshifts, or rule out certain interpretations for single-line detections because of the lack of expected secondary lines.

\subsection{NIRCam WFSS sensitivity}

We also measured the WFSS continuum sensitivity of NEXUS-Wide epoch 1 central mosaics.
We extracted spectra of bright and compact sources, which (1) are brighter than 21 mag in F356W or F444W, (2) have $\geq80\%$ of continuum fluxes concentrated in boxcar-extracted spectra with an aperture height of 5 pixels.
We measure the continuum root-mean-square noise based on the error extension of 1D spectrum.
We also remove the continuum of 1D spectrum through median-filtering (kernel\,=\,0.1\,\micron\,=\,100\,pixels), measure the standard deviation and compare with the median of 1D error spectrum, and thus validating the error spectrum.
Based on the median stacks of 1D error spectra of 78 sources in each band that fall on NIRCam module A, we obtain $3\sigma$ detection limit of continuum flux density (wavelength bin size =\,1\,nm\,=\,1\,pixel) for both filters as shown in Figure~\ref{fig:wfss-sens}.

With the first epoch of observation, NEXUS reaches a $3\sigma$ depth of 21.5\,mag (per pixel) for more than half of the wavelength coverage. 
The most sensitive wavelength is $\sim$3.95\,\micron\ with the F444W band, reaching a $3\sigma$ depth of 21.9\,mag in module A.
We also repeat the experiment for sources fall on NIRCam module B, and the detection limit is shallower by 16\%\ (in $\mu$Jy) because grisms on module B is only anti-reflection coated on the flat side \citep[see][]{Rieke23}.

We also compared with JWST online ETC v4.0\footnote{\href{https://jwst.etc.stsci.edu/}{https://jwst.etc.stsci.edu/}} simulations with similar conditions to our actual observations (e.g., Module A, background on the same observing date and at the same coordinates, point-like source morphology and 5-pixel extraction aperture height). 
The results from ETC are shown as the dashed lines in Figure~\ref{fig:wfss-sens}.
We find that our observations are more sensitive than ETC's prediction at 3.5--5.0\,\micron, which is known as a result of higher grism throughput as found through JWST commissioning \citep{Sun22,Rieke23,Rigby23}.
The F322W2 data are somewhat less sensitive than ETC's prediction at $\lesssim$3.1\,\micron, but the difference is small ($\sim$\,10\%).

The emission-line sensitivity of NEXUS WFSS observations will be presented in the forthcoming DR1 paper, as it requires careful identification of a large number of emission-line galaxies across the full wavelength range.
Preliminary analyses suggest that we have reached a characteristic $5\sigma$ emission line detection limit at $\sim 8\times10^{-18}$ erg s$^{-1}$ cm$^{-2}$ (3.5--4.0\,\micron; F322W2 filter) and $\sim 5\times10^{-18}$ erg s$^{-1}$ cm$^{-2}$ (4.0--4.5\,\micron; F444W filter), respectively.
Similar to that of continuum, the sensitivity decreases towards the blue end of F322W2 filter ($\sim 3\times10^{-17}$ erg s$^{-1}$ cm$^{-2}$ at $\sim$\,2.6\,\micron) and the red end of F444W filter ($\sim 7\times10^{-18}$ erg s$^{-1}$ cm$^{-2}$ at $\sim$\,4.9\,\micron). 
The emission-line sensitivity in the F444W band is better than the pre-observation expectation from ETC \citep{nexus}, and the F322W2 line sensitivity is similar to that of ETC.

\begin{figure*}[ht]
\centering
\includegraphics[width=0.9\textwidth]{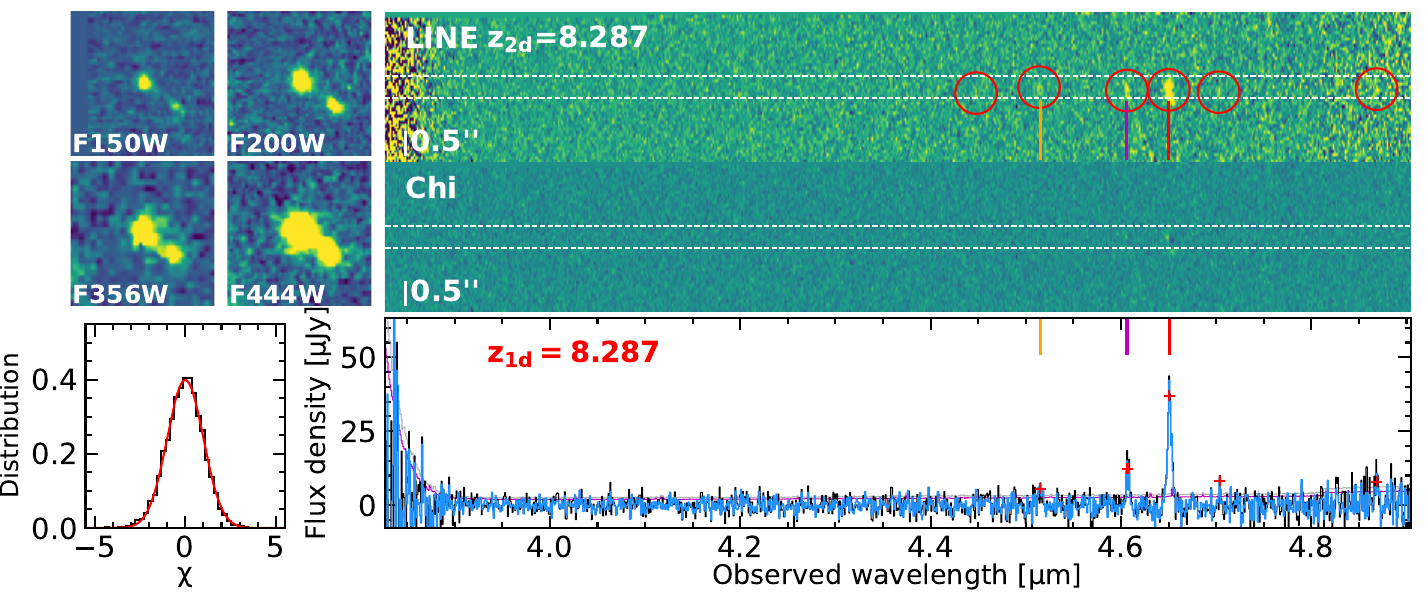}
\caption{
An example of the quality assurance (QA) plot for the WFSS spectra extraction. The image thumbnails show the image cutouts in F150W, F200W, F356W, and F444W filters. The size of the cutouts is $1\farcs5\times1\farcs5$ and the image orientations are rotated to align with the dispersion direction (wavelength increases from left to right) of WFSS image shown in the right. The coadded 2D spectrum (after continuum model subtracted) is shown in the top-right panle and the $\chi$ image where $\rm \chi=\displaystyle\frac{data-optimal~ model}{uncertainty}$ is shown in the middle-right panel. The $\chi$ distribution of all pixels within the extraction aperture (white dotted lines) compared with a Gaussian function with $\sigma=1$ is shown in the bottom left panel. The extracted 1D spectra from the coadded 2D spectrum are shown in the bottom right panel, with blue (magenta) for optimal extraction (error vector), and black (grey) for boxcar extraction (error vector). The red circles in the 2D spectrum plot highlight the potential emission line candidates while the red crosses in the bottom right panel denote those identified from the 1D spectrum. This source is identified as a \OIII-emitting galaxy at $z=8.287$, with the vertical short lines highlight the location of the \OIIIab\ and \hbeta\ lines.
}
\label{fig_wfss_examp}
\end{figure*}

\begin{figure}[!ht]
\centering
\includegraphics[width=0.5\textwidth]{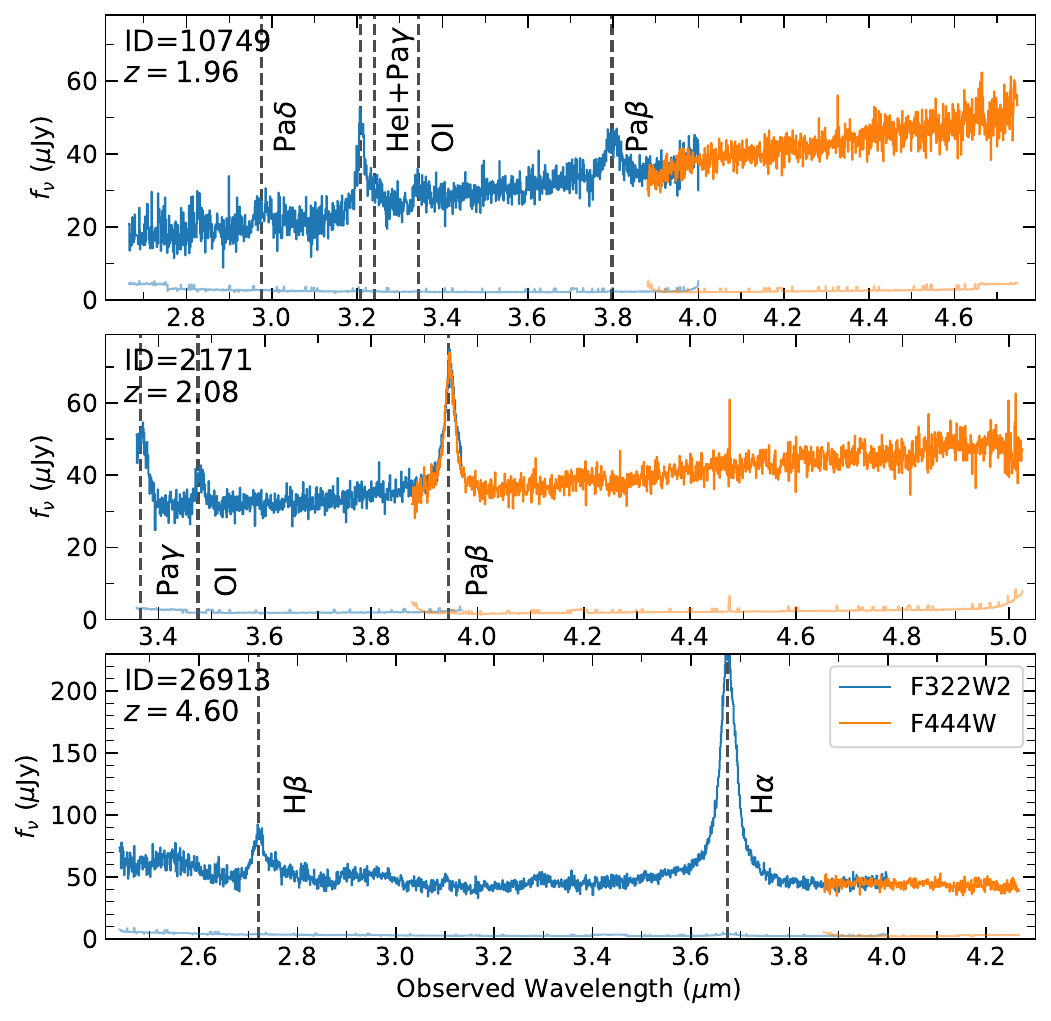}
\caption{WFSS grism spectra of three bright (F444W$<20$ mag) $z=2-5$ quasars with continuum detection from optimal extraction. Blue and orange curves represent F322W2 and F444W grism spectra, with uncertainties shown in fainter colors. Source ID and spectroscopic redshift are shown at the top-left corner. Prominent lines are indicated by vertical dashed lines with labels shown to the right.}
\label{QSO_WFSS}
\end{figure}

\begin{figure}[!ht]
\centering
\includegraphics[width=0.5\textwidth]{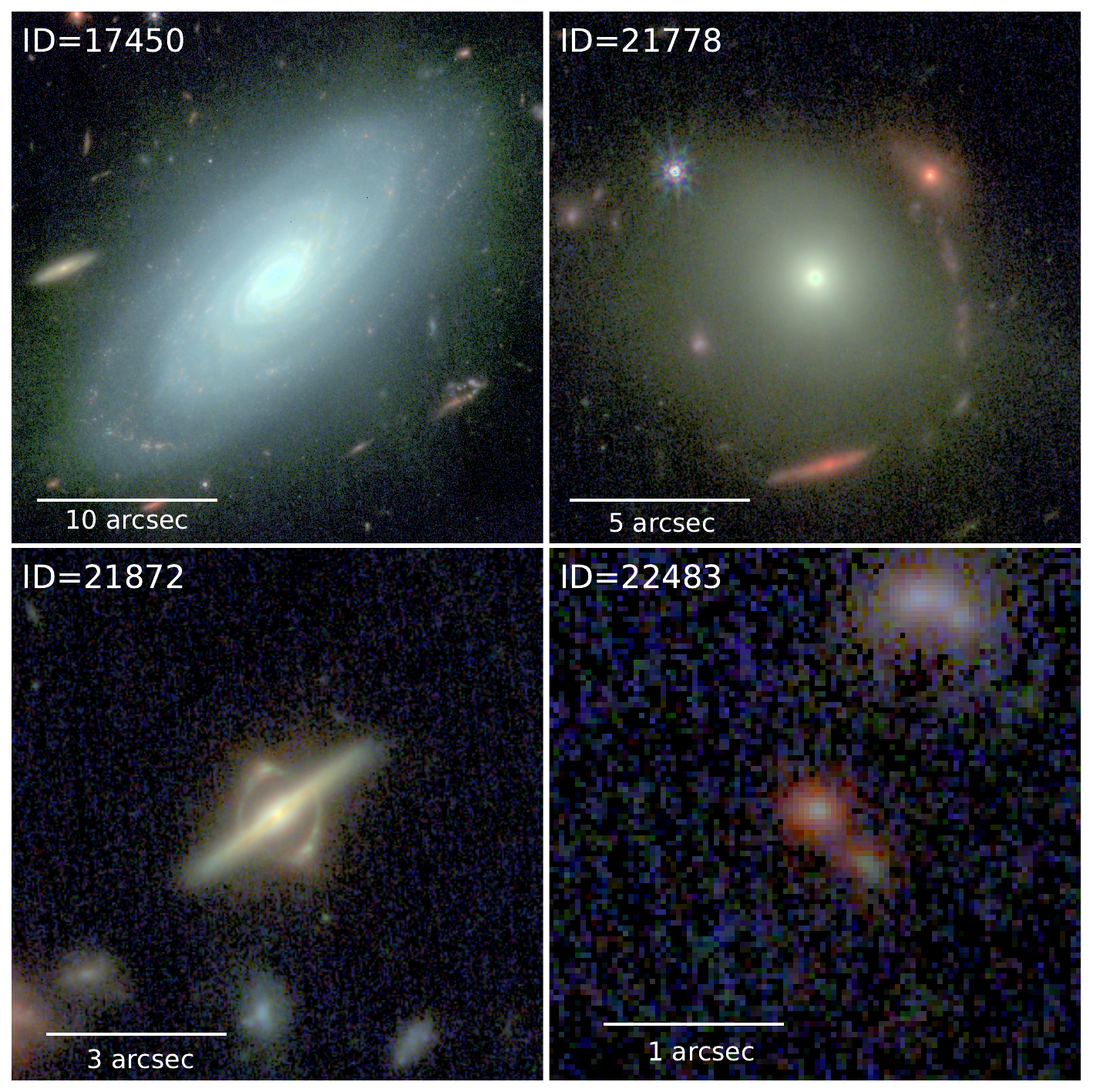}
\caption{Pseudo-color cutouts of four example objects: a local late-type spiral galaxy at $z=0.038$ (top left); a foreground elliptical galaxy at $z=0.544$ with candidate gravitationally-lensed arcs in the background (top right); an edge-on galaxy with central dust lane at $z=0.55$ and an Einstein ring of lensed galaxy at $z=2.61$ (bottom left); a \OIII\ emitting pair at $z=8.287$ (bottom right). Color scheme is the same as Figure~\ref{fig1}. Source ID for the central primary source is shown at the upper-left corner and the scale bar is shown at the lower-left corner.}
\label{IMG_examples}
\end{figure}

\subsection{Illustration of Science Applications}\label{sec:app}

NIRCam WFSS has been widely used for flux-complete selection of high-redshift emission-line galaxies entering its field of view, especially \OIII\ emitters of which the $\lambda\lambda$4959,5007 doublets can be detected at the same time \citep[e.g.,][]{Sun22,Sun23,EIGER,Wang23}.
In Figure~\ref{fig_wfss_examp}, we show a close pair of luminous \OIII-emitting galaxies at $z=8.287$, which are blindly discovered with our NIRCam WFSS data.
Strong \OIIIab\ and weak \hbeta\ emission can be identified in the spectrum.
This source has a total \OIIIb\ luminosity of $(2.35\pm0.13)\times10^{43}$ erg s$^{-1}$, making it one of the most luminous \OIII-emitting galaxies at $z>8$ identified so far (i.e., brighter than any \OIII-emitter identified with the FRESCO survey in the GOODS-S/N; \citealt{FRESCO,Meyer24}). 

Moreover, a combination of F322W2 and F444W WFSS grisms enables a broad wavelength coverage, which is especially useful to select high-$z$ AGNs. Our grisms will cover broad \HeI, Pa$\gamma$, and Pa$\beta$ at $z\approx2$, and broad H$\beta$ and H$\alpha$ at $4\la z \la7$. Figure~\ref{QSO_WFSS} shows NIRCam WFSS spectra of three bright (F444W $<20$ mag) quasars at $2\la z \la5$. Detailed studies of their black hole masses, number density and variability will be the focuses of future papers from the NEXUS program.

The high resolution, multi-band NIRCam imaging allows us to probe galaxies from the local Universe to cosmic dawn. Figure~\ref{IMG_examples} shows pseudo-color cutouts of four example objects. ID17450 is a local late-type spiral galaxy at $z=0.038$. F444W imaging resolves hot dust and polycyclic aromatic hydrocarbons emission from star-forming clumps. ID21778 is a foreground elliptical galaxy at $z=0.544$, with candidate gravitationally-lensed arcs in the background. ID21872 is an edge-on disk galaxy with a central dust lane and a ring like structure. Our preliminary analysis utilizing WFSS grism spectra suggests it is a gravitational lens system, with the foreground lens at $z=0.55$ (confirmed through WFSS detection of CO bandhead absorption at rest-frame 2.3--2.5\,\micron) and the background galaxy at $z=2.61$ (confirmed through WFSS detections of \SIII\,$\lambda$9531 and \HeI\,$\lambda$10830) lensed as an Einstein ring. ID22483 refers to the source ID of the brighter member of the \OIII\ emitting pair at $z=8.287$, whose F444W grism spectrum is shown in Figure~\ref{fig_wfss_examp}.

\section{EDR Data Products, Access, and Caveats}\label{sec6}

In this paper, we release the fully reduced images, photometric catalog, and stacked 1D NIRCam F322W2 and F444W2 WFSS spectra of bright (F356W $<21$ mag or F444W $<21$ mag) objects. Table~\ref{table2} shows the format of our photometric catalog. The photometric catalog contains source coordinates, sizes, aperture fluxes in six apertures ($r=0\farcs15, 0\farcs25, 0\farcs3, 0\farcs35, 0\farcs5$), AUTO fluxes within Kron aperture, six total fluxes based on six aperture fluxes after aperture correction, flux use flag, \texttt{SExtractor} source extraction flag, and \texttt{SExtractor} star-galaxy classifier. For image mosaics, we release data, error, and weight extensions as separate fits files in each of the six filters. For the 874 sources with EDR WFSS spectra, we include both optimal- and boxcar-extracted spectra, with each of them having two versions: before and after continuum subtraction. Data products are publicly accessible at \url{https://ariel.astro.illinois.edu/nexus/edr/}. We also provide an online interactive map for quick visualization of released images and WFSS spectra at \url{https://ariel.astro.illinois.edu/nexus/map/} using \texttt{FitsMap} \citep{hausen2022a}.

According to our survey design, individual epoch observations do not have full area or filter coverage. In this release, F150W and F356W covers $\sim$83\% while F115W and F200W covers $\sim$93\% of FoV in the F090W and F444W filters, leaving a significant amount of sources lacking complete photometry. There are also some very bright stars in the NEXUS field, which are saturated in most of the filters. The coordinates of the these sources may have some small offsets. Their bright spikes may also degrade the performance of deblending and photometry of nearby sources. We also notice low level ``claw'' features around the corner of the A4 detector in a few F200W exposures. No source photometry is affected by these ``claw'' features. We will be able to resolve most of these issues in DR1 by combining NEXUS-Wide epoch 2 and NEXUS-Deep exposures.

As mentioned in Section \ref{wfss_reduction}, the stacked 1D spectra have better outlier rejection and object profile fitting for bright continuum sources. However, one caveat of slitless observations is that they suffer from significant spectral contamination issues (i.e., a given spectrum can be contaminated by emission from other sources along the dispersion direction). Caution is advised when using the spectra for very extended sources, as the aperture used for boxcar extraction and the object profile may not be optimal. To help users assess the contamination issue and other potential problems with the WFSS spectra, we have included a visually inspected flag for each source {(summarized in the file \texttt{wide\_ep1\_01\_wfss\_bright\_src\_flag.fits} in the EDR repository)}.
The flags indicate sources with good continuum and/or line detections, or subject to the contamination of other bright continuum sources. The visual inspection is by no means complete, and will be updated in future data releases.

Higher-order data products, including spectroscopic redshifts, photometric redshifts (phot-zs), emission line fluxes, and line-emitter catalogs will be presented in the forthcoming DR1. 

\begin{deluxetable*}{llcl}[ht]
\caption{FITS table format for the NEXUS-Wide Epoch1 photometric source catalog \label{table2}}
\tablehead{
\colhead{Column Name} & \colhead{Format} & \colhead{Units} & \colhead{Description}
}
\startdata
\texttt{ID} & LONG &  & Source ID\\
\texttt{RA} & DOUBLE & degree & Right Ascension (J2000) \\
\texttt{DEC} & DOUBLE & degree & Declination (J2000) \\
\texttt{RA\_WIN} & DOUBLE & degree & Windowed Right Ascension\\
\texttt{DEC\_WIN} & DOUBLE & degree & Windowed Declination\\
\texttt{R\_KRON} & DOUBLE & arcsec & Kron radius\\
\texttt{A\_IMAGE} & DOUBLE & pix & Source elliptical semi-major axis\\
\texttt{B\_IMAGE} & DOUBLE & pix & Source elliptical semi-minor axis\\
\texttt{THETA\_IMAGE} & DOUBLE & deg & Position angle of \texttt{A\_IMAGE} from positive X-axis counter-clockwise\\
\texttt{FLUX\_AUTO\_[band]} & DOUBLE & $\mu$Jy & Kron flux in \texttt{band} after aperture flux loss correction\\
\texttt{FLUXERR\_AUTO\_[band]} & DOUBLE & $\mu$Jy & Kron flux error\\
\texttt{MEAN\_NEXP\_AUTO\_APER\_[band]} & SHORT &  & Mean number of exposure within Kron aperture\\
\texttt{MASK\_FLUX\_AUTO\_[band]} & BOOL &  & If \texttt{FLUX\_AUTO\_[band]} and its error are masked \\
\texttt{FLUX\_APER\_[band]} & DOUBLE & $\mu$Jy & Circular aperture flux within r=0\farcs1\\
\texttt{FLUXERR\_APER\_[band]} & DOUBLE & $\mu$Jy & Error of \texttt{FLUX\_APER\_[band]}\\
\texttt{FLUX\_TOTAL\_APER\_[band]} & DOUBLE & $\mu$Jy & Total flux based on \texttt{FLUX\_APER\_[band]}\\
\texttt{FLUXERR\_TOTAL\_APER\_[band]} & DOUBLE & $\mu$Jy & Total flux error based on \texttt{FLUXERR\_APER\_[band]}\\
\texttt{MEAN\_NEXP\_FIXED\_APER\_[band]} & SHORT &  & Mean number of exposure within r=0\farcs1 circular aperture\\
\texttt{MASK\_FLUX\_FIXED\_APER\_[band]} & BOOL &  & If \texttt{FLUX\_APER\_[band]} and its error are masked \\
\texttt{FLUX\_APER\_[1-5]\_[band]} & DOUBLE & $\mu$Jy & Circular aperture flux within \texttt{APER\_[1-5]}\\
\texttt{FLUXERR\_APER\_[1-5]\_[band]} & DOUBLE & $\mu$Jy & Error of \texttt{FLUX\_FIXED\_APER\_[band]}\\
\texttt{FLUX\_TOTAL\_APER\_[1-5]\_[band]} & DOUBLE & $\mu$Jy & Total flux based on \texttt{FLUX\_APER\_[1-5]\_[band]}\\
\texttt{FLUXERR\_TOTAL\_APER\_[1-5]\_[band]} & DOUBLE & $\mu$Jy & Total flux error based on \texttt{FLUXERR\_APER\_[1-5]\_[band]}\\
\texttt{MEAN\_NEXP\_FIXED\_APER\_[1-5]\_[band]} & LONG &  & Mean number of exposure for \texttt{APER\_[1-5]}\\
\texttt{MASK\_FLUX\_FIXED\_APER\_[1-5]\_[band]} & BOOL &  & If \texttt{FLUX\_APER\_[1-5]\_[band]} and its error are masked \\
\texttt{FLUX\_RADIUS\_[band]} & DOUBLE & arcsec & Half light radius within \texttt{FLUX\_AUTO\_[band]}\\
\texttt{FLAGS\_[band]} & LONG &  & \texttt{SExtractor} extraction flags\\
\texttt{CLASS\_STAR\_[band]} & DOUBLE & & \texttt{SExtractor} neural network-based star-galaxy classifier\\
\texttt{F\_TOTAL\_APER} & DOUBLE & & Scaling factor from \texttt{FLUX\_APER\_[band]} to \texttt{FLUX\_TOTAL\_APER\_[band]}  \\
\texttt{F\_TOTAL\_APER\_[1-5]} & DOUBLE & & Scaling factor from \texttt{FLUX\_APER\_[1-5]\_[band]} to \\
&&&\texttt{FLUX\_TOTAL\_APER\_[1-5]\_[band]}\\
\enddata
\tablecomments{ \texttt{band} represents the F090W, F115W, F150W, F200W, F356W, and F444W filters. \texttt{APER\_[1-5]} are circular apertures with $r=0\farcs15, 0\farcs25, 0\farcs3, 0\farcs35, 0\farcs5$, respectively. We use $-99$ to indicate missing data due to either out-of-coverage or bad pixels (\texttt{MASK\_FLUX\_*=True}).}
\end{deluxetable*}

\begin{acknowledgments}

Based on observations with the NASA/ESA/CSA James Webb Space Telescope obtained from the Barbara A. Mikulski Archive at the Space Telescope Science Institute, which is operated by the Association of Universities for Research in Astronomy, Incorporated, under NASA contract NAS5-03127. Support for Program numbers JWST-GO-02057, JWST-AR-03038, and JWST-GO-05105 (YS, MZ and JL) was provided through a grant from the STScI under NASA contract NAS5-03127. GN gratefully acknowledges NSF CAREER grant AST-2239364, supported in-part by funding from Charles Simonyi, support from NSF AST-2206195, and NSF OAC-2311355, DOE support through the Department of Physics at the University of Illinois, Urbana-Champaign (13771275), and support from the HST Guest Observer Program through HST-GO-16764 and HST-GO-17128 (PI: R. Foley).
\end{acknowledgments}

%

\vspace{5mm}
\facilities{JWST (NIRCam)}


\software{
\texttt{Astropy} \citep{2013A&A...558A..33A,2018AJ....156..123A}, 
\texttt{FitsMap} \citep{hausen2022a},
\texttt{jwst} \url{https://jwst-pipeline.readthedocs.io/en/latest/},
\texttt{Matplotlib} \citep{Hunter2007}, 
\texttt{Numpy} \citep{Harris2020}, 
\texttt{photutils} \citep{photutils}, 
\texttt{PSFEx} \citep{PSFEx}, 
\texttt{scipy} \citep{scipy}, 
\texttt{Source Extractor} \citep{1996A&AS..117..393B}
          }





\bibliography{sample631}{}

\begin{thebibliography}{}
\expandafter\ifx\csname natexlab\endcsname\relax\def\natexlab#1{#1}\fi
\providecommand{\url}[1]{\href{#1}{#1}}
\providecommand{\dodoi}[1]{doi:~\href{http://doi.org/#1}{\nolinkurl{#1}}}
\providecommand{\doeprint}[1]{\href{http://ascl.net/#1}{\nolinkurl{http://ascl.net/#1}}}
\providecommand{\doarXiv}[1]{\href{https://arxiv.org/abs/#1}{\nolinkurl{https://arxiv.org/abs/#1}}}

\bibitem[{{Astropy Collaboration} {et~al.}(2013){Astropy Collaboration},
  {Robitaille}, {Tollerud}, {Greenfield}, {Droettboom}, {Bray}, {Aldcroft},
  {Davis}, {Ginsburg}, {Price-Whelan}, {Kerzendorf}, {Conley}, {Crighton},
  {Barbary}, {Muna}, {Ferguson}, {Grollier}, {Parikh}, {Nair}, {Unther},
  {Deil}, {Woillez}, {Conseil}, {Kramer}, {Turner}, {Singer}, {Fox}, {Weaver},
  {Zabalza}, {Edwards}, {Azalee Bostroem}, {Burke}, {Casey}, {Crawford},
  {Dencheva}, {Ely}, {Jenness}, {Labrie}, {Lim}, {Pierfederici}, {Pontzen},
  {Ptak}, {Refsdal}, {Servillat}, \& {Streicher}}]{2013A&A...558A..33A}
{Astropy Collaboration}, {Robitaille}, T.~P., {Tollerud}, E.~J., {et~al.} 2013,
  \aap, 558, A33, \dodoi{10.1051/0004-6361/201322068}

\bibitem[{{Astropy Collaboration} {et~al.}(2018){Astropy Collaboration},
  {Price-Whelan}, {Sip{\H{o}}cz}, {G{\"u}nther}, {Lim}, {Crawford}, {Conseil},
  {Shupe}, {Craig}, {Dencheva}, {Ginsburg}, {VanderPlas}, {Bradley},
  {P{\'e}rez-Su{\'a}rez}, {de Val-Borro}, {Aldcroft}, {Cruz}, {Robitaille},
  {Tollerud}, {Ardelean}, {Babej}, {Bach}, {Bachetti}, {Bakanov}, {Bamford},
  {Barentsen}, {Barmby}, {Baumbach}, {Berry}, {Biscani}, {Boquien}, {Bostroem},
  {Bouma}, {Brammer}, {Bray}, {Breytenbach}, {Buddelmeijer}, {Burke},
  {Calderone}, {Cano Rodr{\'\i}guez}, {Cara}, {Cardoso}, {Cheedella}, {Copin},
  {Corrales}, {Crichton}, {D'Avella}, {Deil}, {Depagne}, {Dietrich}, {Donath},
  {Droettboom}, {Earl}, {Erben}, {Fabbro}, {Ferreira}, {Finethy}, {Fox},
  {Garrison}, {Gibbons}, {Goldstein}, {Gommers}, {Greco}, {Greenfield},
  {Groener}, {Grollier}, {Hagen}, {Hirst}, {Homeier}, {Horton}, {Hosseinzadeh},
  {Hu}, {Hunkeler}, {Ivezi{\'c}}, {Jain}, {Jenness}, {Kanarek}, {Kendrew},
  {Kern}, {Kerzendorf}, {Khvalko}, {King}, {Kirkby}, {Kulkarni}, {Kumar},
  {Lee}, {Lenz}, {Littlefair}, {Ma}, {Macleod}, {Mastropietro}, {McCully},
  {Montagnac}, {Morris}, {Mueller}, {Mumford}, {Muna}, {Murphy}, {Nelson},
  {Nguyen}, {Ninan}, {N{\"o}the}, {Ogaz}, {Oh}, {Parejko}, {Parley}, {Pascual},
  {Patil}, {Patil}, {Plunkett}, {Prochaska}, {Rastogi}, {Reddy Janga},
  {Sabater}, {Sakurikar}, {Seifert}, {Sherbert}, {Sherwood-Taylor}, {Shih},
  {Sick}, {Silbiger}, {Singanamalla}, {Singer}, {Sladen}, {Sooley},
  {Sornarajah}, {Streicher}, {Teuben}, {Thomas}, {Tremblay}, {Turner},
  {Terr{\'o}n}, {van Kerkwijk}, {de la Vega}, {Watkins}, {Weaver}, {Whitmore},
  {Woillez}, {Zabalza}, \& {Astropy Contributors}}]{2018AJ....156..123A}
{Astropy Collaboration}, {Price-Whelan}, A.~M., {Sip{\H{o}}cz}, B.~M., {et~al.}
  2018, \aj, 156, 123, \dodoi{10.3847/1538-3881/aabc4f}

\bibitem[{{Bagley} {et~al.}(2023){Bagley}, {Pirzkal}, {Finkelstein},
  {Papovich}, {Berg}, {Lotz}, {Leung}, {Ferguson}, {Koekemoer}, {Dickinson},
  {Kartaltepe}, {Kocevski}, {Somerville}, {Yung}, {Backhaus}, {Casey},
  {Castellano}, {Ch{\'a}vez Ortiz}, {Chworowsky}, {Cox}, {Dav{\'e}}, {Davis},
  {Estrada-Carpenter}, {Fontana}, {Fujimoto}, {Gardner}, {Giavalisco},
  {Grazian}, {Grogin}, {Hathi}, {Hutchison}, {Jaskot}, {Jung}, {Kewley},
  {Kirkpatrick}, {Larson}, {Matharu}, {Natarajan}, {Pentericci},
  {P{\'e}rez-Gonz{\'a}lez}, {Ravindranath}, {Rothberg}, {Ryan}, {Shen},
  {Simons}, {Snyder}, {Trump}, \& {Wilkins}}]{NGDEEP}
{Bagley}, M.~B., {Pirzkal}, N., {Finkelstein}, S.~L., {et~al.} 2023, arXiv
  e-prints, arXiv:2302.05466, \dodoi{10.48550/arXiv.2302.05466}

\bibitem[{{Bertin}(2011)}]{PSFEx}
{Bertin}, E. 2011, in Astronomical Society of the Pacific Conference Series,
  Vol. 442, Astronomical Data Analysis Software and Systems XX, ed. I.~N.
  {Evans}, A.~{Accomazzi}, D.~J. {Mink}, \& A.~H. {Rots}, 435

\bibitem[{{Bertin} \& {Arnouts}(1996)}]{1996A&AS..117..393B}
{Bertin}, E., \& {Arnouts}, S. 1996, \aaps, 117, 393,
  \dodoi{10.1051/aas:1996164}

\bibitem[{{Bezanson} {et~al.}(2022){Bezanson}, {Labbe}, {Whitaker}, {Leja},
  {Price}, {Franx}, {Brammer}, {Marchesini}, {Zitrin}, {Wang}, {Weaver},
  {Furtak}, {Atek}, {Coe}, {Cutler}, {Dayal}, {van Dokkum}, {Feldmann},
  {Forster Schreiber}, {Fujimoto}, {Geha}, {Glazebrook}, {de Graaff}, {Greene},
  {Juneau}, {Kassin}, {Kriek}, {Khullar}, {Maseda}, {Mowla}, {Muzzin},
  {Nanayakkara}, {Nelson}, {Oesch}, {Pacifici}, {Pan}, {Papovich}, {Setton},
  {Shapley}, {Smit}, {Stefanon}, {Taylor}, \& {Williams}}]{UNCOVER}
{Bezanson}, R., {Labbe}, I., {Whitaker}, K.~E., {et~al.} 2022, arXiv e-prints,
  arXiv:2212.04026, \dodoi{10.48550/arXiv.2212.04026}

\bibitem[{{Boucaud} {et~al.}(2016){Boucaud}, {Bocchio}, {Abergel}, {Orieux},
  {Dole}, \& {Hadj-Youcef}}]{pypher}
{Boucaud}, A., {Bocchio}, M., {Abergel}, A., {et~al.} 2016, \aap, 596, A63,
  \dodoi{10.1051/0004-6361/201629080}

\bibitem[{Bradley {et~al.}(2022)Bradley, Sip{\H o}cz, Robitaille, Tollerud,
  Vin{\'{\i}}cius, Deil, Barbary, Wilson, Busko, G{\"u}nther, Cara, Conseil,
  Bostroem, Droettboom, Bray, Bratholm, Lim, Barentsen, Craig, Pascual, Perren,
  Greco, Donath, de~Val-Borro, Kerzendorf, Bach, Weaver, D'Eugenio, Souchereau,
  \& Ferreira}]{photutils}
Bradley, L., Sip{\H o}cz, B., Robitaille, T., {et~al.} 2022, astropy/photutils:
  1.5.0, 1.5.0,  Zenodo, \dodoi{10.5281/zenodo.6825092}

\bibitem[{Bushouse {et~al.}(2023)Bushouse, Eisenhamer, Dencheva, Davies,
  Greenfield, Morrison, Hodge, Simon, Grumm, Droettboom, Slavich, Sosey, Pauly,
  Miller, Jedrzejewski, Hack, Davis, Crawford, Law, Gordon, Regan, Cara,
  MacDonald, Bradley, Shanahan, Jamieson, Teodoro, Williams, Pena-Guerrero, \&
  Graham}]{Bushouse2023}
Bushouse, H., Eisenhamer, J., Dencheva, N., {et~al.} 2023, {JWST Calibration
  Pipeline}, 1.10.2, \dodoi{10.5281/zenodo.7829329}

\bibitem[{Bushouse {et~al.}(2024)Bushouse, Eisenhamer, Dencheva, Davies,
  Greenfield, Morrison, Hodge, Simon, Grumm, Droettboom, Slavich, Sosey, Pauly,
  Miller, Jedrzejewski, Hack, Davis, Crawford, Law, Gordon, Regan, Cara,
  MacDonald, Bradley, Shanahan, Jamieson, Teodoro, Williams, Pena-Guerrero, \&
  Graham}]{Bushouse_JWST_Calibration_Pipeline_2024}
---. 2024, {JWST Calibration Pipeline}, 1.15.1, \dodoi{10.5281/zenodo.12692459}

\bibitem[{{Carnall} {et~al.}(2023){Carnall}, {McLure}, {Dunlop}, {McLeod},
  {Wild}, {Cullen}, {Magee}, {Begley}, {Cimatti}, {Donnan}, {Hamadouche},
  {Jewell}, \& {Walker}}]{2023Natur.619..716C}
{Carnall}, A.~C., {McLure}, R.~J., {Dunlop}, J.~S., {et~al.} 2023, \nat, 619,
  716, \dodoi{10.1038/s41586-023-06158-6}

\bibitem[{{Carnall} {et~al.}(2024){Carnall}, {Cullen}, {McLure}, {McLeod},
  {Begley}, {Donnan}, {Dunlop}, {Shapley}, {Rowlands}, {Almaini},
  {Arellano-C{\'o}rdova}, {Barrufet}, {Cimatti}, {Ellis}, {Grogin},
  {Hamadouche}, {Illingworth}, {Koekemoer}, {Leung}, {Lovell},
  {P{\'e}rez-Gonz{\'a}lez}, {Santini}, {Stanton}, \&
  {Wild}}]{2024MNRAS.534..325C}
{Carnall}, A.~C., {Cullen}, F., {McLure}, R.~J., {et~al.} 2024, \mnras, 534,
  325, \dodoi{10.1093/mnras/stae2092}

\bibitem[{{Carniani} {et~al.}(2024){Carniani}, {Hainline}, {D'Eugenio},
  {Eisenstein}, {Jakobsen}, {Witstok}, {Johnson}, {Chevallard}, {Maiolino},
  {Helton}, {Willott}, {Robertson}, {Alberts}, {Arribas}, {Baker},
  {Bhatawdekar}, {Boyett}, {Bunker}, {Cameron}, {Cargile}, {Charlot}, {Curti},
  {Curtis-Lake}, {Egami}, {Giardino}, {Isaak}, {Ji}, {Jones}, {Kumari},
  {Maseda}, {Parlanti}, {P{\'e}rez-Gonz{\'a}lez}, {Rawle}, {Rieke}, {Rieke},
  {Del Pino}, {Saxena}, {Scholtz}, {Smit}, {Sun}, {Tacchella}, {{\"U}bler},
  {Venturi}, {Williams}, \& {Willmer}}]{2024Natur.633..318C}
{Carniani}, S., {Hainline}, K., {D'Eugenio}, F., {et~al.} 2024, \nat, 633, 318,
  \dodoi{10.1038/s41586-024-07860-9}

\bibitem[{{Casertano} {et~al.}(2000){Casertano}, {de Mello}, {Dickinson},
  {Ferguson}, {Fruchter}, {Gonzalez-Lopezlira}, {Heyer}, {Hook}, {Levay},
  {Lucas}, {Mack}, {Makidon}, {Mutchler}, {Smith}, {Stiavelli}, {Wiggs}, \&
  {Williams}}]{2000AJ....120.2747C}
{Casertano}, S., {de Mello}, D., {Dickinson}, M., {et~al.} 2000, \aj, 120,
  2747, \dodoi{10.1086/316851}

\bibitem[{{Casey} {et~al.}(2023){Casey}, {Kartaltepe}, {Drakos}, {Franco},
  {Harish}, {Paquereau}, {Ilbert}, {Rose}, {Cox}, {Nightingale}, {Robertson},
  {Silverman}, {Koekemoer}, {Massey}, {McCracken}, {Rhodes}, {Akins}, {Allen},
  {Amvrosiadis}, {Arango-Toro}, {Bagley}, {Bongiorno}, {Capak}, {Champagne},
  {Chartab}, {Ch{\'a}vez Ortiz}, {Chworowsky}, {Cooke}, {Cooper}, {Darvish},
  {Ding}, {Faisst}, {Finkelstein}, {Fujimoto}, {Gentile}, {Gillman}, {Gould},
  {Gozaliasl}, {Hayward}, {He}, {Hemmati}, {Hirschmann}, {Jahnke}, {Jin},
  {Khostovan}, {Kokorev}, {Lambrides}, {Laigle}, {Larson}, {Leung}, {Liu},
  {Liaudat}, {Long}, {Magdis}, {Mahler}, {Mainieri}, {Manning}, {Maraston},
  {Martin}, {McCleary}, {McKinney}, {McPartland}, {Mobasher}, {Pattnaik},
  {Renzini}, {Rich}, {Sanders}, {Sattari}, {Scognamiglio}, {Scoville}, {Sheth},
  {Shuntov}, {Sparre}, {Suzuki}, {Talia}, {Toft}, {Trakhtenbrot}, {Urry},
  {Valentino}, {Vanderhoof}, {Vardoulaki}, {Weaver}, {Whitaker}, {Wilkins},
  {Yang}, \& {Zavala}}]{COSMOS-Web}
{Casey}, C.~M., {Kartaltepe}, J.~S., {Drakos}, N.~E., {et~al.} 2023, \apj, 954,
  31, \dodoi{10.3847/1538-4357/acc2bc}

\bibitem[{{Curtis-Lake} {et~al.}(2023){Curtis-Lake}, {Carniani}, {Cameron},
  {Charlot}, {Jakobsen}, {Maiolino}, {Bunker}, {Witstok}, {Smit}, {Chevallard},
  {Willott}, {Ferruit}, {Arribas}, {Bonaventura}, {Curti}, {D'Eugenio},
  {Franx}, {Giardino}, {Looser}, {L{\"u}tzgendorf}, {Maseda}, {Rawle}, {Rix},
  {Rodr{\'\i}guez del Pino}, {{\"U}bler}, {Sirianni}, {Dressler}, {Egami},
  {Eisenstein}, {Endsley}, {Hainline}, {Hausen}, {Johnson}, {Rieke},
  {Robertson}, {Shivaei}, {Stark}, {Tacchella}, {Williams}, {Willmer},
  {Bhatawdekar}, {Bowler}, {Boyett}, {Chen}, {de Graaff}, {Helton}, {Hviding},
  {Jones}, {Kumari}, {Lyu}, {Nelson}, {Perna}, {Sandles}, {Saxena}, {Suess},
  {Sun}, {Topping}, {Wallace}, \& {Whitler}}]{2023NatAs...7..622C}
{Curtis-Lake}, E., {Carniani}, S., {Cameron}, A., {et~al.} 2023, Nature
  Astronomy, 7, 622, \dodoi{10.1038/s41550-023-01918-w}

\bibitem[{{de Graaff} {et~al.}(2024){de Graaff}, {Setton}, {Brammer}, {Cutler},
  {Suess}, {Labbe}, {Leja}, {Weibel}, {Maseda}, {Whitaker}, {Bezanson},
  {Boogaard}, {Cleri}, {De Lucia}, {Franx}, {Greene}, {Hirschmann}, {Matthee},
  {McConachie}, {Naidu}, {Oesch}, {Price}, {Rix}, {Valentino}, {Wang}, \&
  {Williams}}]{2024arXiv240405683D}
{de Graaff}, A., {Setton}, D.~J., {Brammer}, G., {et~al.} 2024, arXiv e-prints,
  arXiv:2404.05683, \dodoi{10.48550/arXiv.2404.05683}

\bibitem[{{Dunlop} {et~al.}(2021){Dunlop}, {Abraham}, {Ashby}, {Bagley},
  {Best}, {Bongiorno}, {Bouwens}, {Bowler}, {Brammer}, {Bremer}, {Calabro'},
  {Carnall}, {Castellano}, {Cirasuolo}, {Conselice}, {Cullen}, {Dave}, {Dayal},
  {Dekel}, {Dickinson}, {Duncan}, {Elbaz}, {Ellis}, {Ferguson}, {Ferrara},
  {Finkelstein}, {Fontana}, {Furlanetto}, {Fynbo}, {Gallerani}, {Gardner},
  {Giavalisco}, {Grazian}, {Grogin}, {Harikane}, {Hopkins}, {Ilbert},
  {Illingworth}, {Juneau}, {Jung}, {Kartaltepe}, {Kassin}, {Kauffmann},
  {Khochfar}, {Kirkpatrick}, {Kocevski}, {Koekemoer}, {Labbe}, {Laporte},
  {Larson}, {Lucas}, {Magee}, {Mason}, {McCracken}, {McLeod}, {McLure},
  {Merlin}, {Mesinger}, {Milvang-Jensen}, {Newman}, {Oesch}, {Ouchi},
  {Pacifici}, {Papovich}, {Peacock}, {Peeples}, {Pentericci}, {Perez-Gonzalez},
  {Pirzkal}, {Pope}, {Pye}, {Reddy}, {Robertson}, {Salvato}, {Santini},
  {Schaerer}, {Shapley}, {Simons}, {Smit}, {Smith}, {Snyder}, {Somerville},
  {Stanway}, {Stefanon}, {Tasca}, {Tikkanen}, {Tresse}, {Trump}, {Whitaker},
  {Wilkins}, {Wright}, {Wyithe}, {van Dokkum}, \& {van der Werf}}]{PRIMER}
{Dunlop}, J.~S., {Abraham}, R.~G., {Ashby}, M. L.~N., {et~al.} 2021, {PRIMER:
  Public Release IMaging for Extragalactic Research}, JWST Proposal. Cycle 1,
  ID. \#1837

\bibitem[{{Eisenstein} {et~al.}(2023){Eisenstein}, {Willott}, {Alberts},
  {Arribas}, {Bonaventura}, {Bunker}, {Cameron}, {Carniani}, {Charlot},
  {Curtis-Lake}, {D'Eugenio}, {Endsley}, {Ferruit}, {Giardino}, {Hainline},
  {Hausen}, {Jakobsen}, {Johnson}, {Maiolino}, {Rieke}, {Rieke}, {Rix},
  {Robertson}, {Stark}, {Tacchella}, {Williams}, {Willmer}, {Baker}, {Baum},
  {Bhatawdekar}, {Boyett}, {Chen}, {Chevallard}, {Circosta}, {Curti},
  {Danhaive}, {DeCoursey}, {de Graaff}, {Dressler}, {Egami}, {Helton},
  {Hviding}, {Ji}, {Jones}, {Kumari}, {L{\"u}tzgendorf}, {Laseter}, {Looser},
  {Lyu}, {Maseda}, {Nelson}, {Parlanti}, {Perna}, {Pusk{\'a}s}, {Rawle},
  {Rodr{\'\i}guez Del Pino}, {Sandles}, {Saxena}, {Scholtz}, {Sharpe},
  {Shivaei}, {Silcock}, {Simmonds}, {Skarbinski}, {Smit}, {Stone}, {Suess},
  {Sun}, {Tang}, {Topping}, {{\"U}bler}, {Villanueva}, {Wallace}, {Whitler},
  {Witstok}, \& {Woodrum}}]{JADES}
{Eisenstein}, D.~J., {Willott}, C., {Alberts}, S., {et~al.} 2023, arXiv
  e-prints, arXiv:2306.02465, \dodoi{10.48550/arXiv.2306.02465}

\bibitem[{{Euclid Collaboration} {et~al.}(2024){Euclid Collaboration},
  {Mellier}, {Abdurro'uf}, {Acevedo Barroso}, {Ach{\'u}carro}, {Adamek},
  {Adam}, {Addison}, {Aghanim}, {Aguena}, {Ajani}, {Akrami}, {Al-Bahlawan},
  {Alavi}, {Albuquerque}, {Alestas}, {Alguero}, {Allaoui}, {Allen}, {Allevato},
  {Alonso-Tetilla}, {Altieri}, {Alvarez-Candal}, {Alvi}, {Amara}, {Amendola},
  {Amiaux}, {Andika}, {Andreon}, {Andrews}, {Angora}, {Angulo}, {Annibali},
  {Anselmi}, {Anselmi}, {Arcari}, {Archidiacono}, {Aric{\`o}}, {Arnaud},
  {Arnouts}, {Asgari}, {Asorey}, {Atayde}, {Atek}, {Atrio-Barandela}, {Aubert},
  {Aubourg}, {Auphan}, {Auricchio}, {Aussel}, {Aussel}, {Avelino},
  {Avgoustidis}, {Avila}, {Awan}, {Azzollini}, {Baccigalupi}, {Bachelet},
  {Bacon}, {Baes}, {Bagley}, {Bahr-Kalus}, {Balaguera-Antolinez}, {Balbinot},
  {Balcells}, {Baldi}, {Baldry}, {Balestra}, {Ballardini}, {Ballester},
  {Balogh}, {Ba{\~n}ados}, {Barbier}, {Bardelli}, {Baron}, {Barreiro},
  {Barrena}, {Barriere}, {Barros}, {Barthelemy}, {Bartolo}, {Basset},
  {Battaglia}, {Battisti}, {Baugh}, {Baumont}, {Bazzanini}, {Beaulieu},
  {Beckmann}, {Belikov}, {Bel}, {Bellagamba}, {Bella}, {Bellini}, {Benabed},
  {Bender}, {Benevento}, {Bennett}, {Benson}, {Bergamini}, {Bermejo-Climent},
  {Bernardeau}, {Bertacca}, {Berthe}, {Berthier}, {Bethermin}, {Beutler},
  {Bevillon}, {Bhargava}, {Bhatawdekar}, {Bianchi}, {Bisigello}, {Biviano},
  {Blake}, {Blanchard}, {Blazek}, {Blot}, {Bosco}, {Bodendorf}, {Boenke},
  {B{\"o}hringer}, {Boldrini}, {Bolzonella}, {Bonchi}, {Bonici}, {Bonino},
  {Bonino}, {Bonvin}, {Bon}, {Booth}, {Borgani}, {Borlaff}, {Borsato}, {Bosco},
  {Bose}, {Botticella}, {Boucaud}, {Bouche}, {Boucher}, {Boutigny}, {Bouvard},
  {Bouwens}, {Bouy}, {Bowler}, {Bozza}, {Bozzo}, {Branchini}, {Brando},
  {Brau-Nogue}, {Brekke}, {Bremer}, {Brescia}, {Breton}, {Brinchmann},
  {Brinckmann}, {Brockley-Blatt}, {Brodwin}, {Brouard}, {Brown}, {Bruton},
  {Bucko}, {Buddelmeijer}, {Buenadicha}, {Buitrago}, {Burger}, {Burigana},
  {Busillo}, {Busonero}, {Cabanac}, {Cabayol-Garcia}, {Cagliari}, {Caillat},
  {Caillat}, {Calabrese}, {Calabro}, {Calderone}, {Calura}, {Camacho Quevedo},
  {Camera}, {Campos}, {Canas-Herrera}, {Candini}, {Cantiello}, {Capobianco},
  {Cappellaro}, {Cappelluti}, {Cappi}, {Caputi}, {Cara}, {Carbone}, {Cardone},
  {Carella}, {Carlberg}, {Carle}, {Carminati}, {Caro}, {Carrasco}, {Carretero},
  {Carrilho}, {Carron Duque}, {Carry}, {Carvalho}, {Carvalho}, {Casas},
  {Casas}, {Casenove}, {Casey}, {Cassata}, {Castander}, {Castelao},
  {Castellano}, {Castiblanco}, {Castignani}, {Castro}, {Cavet}, {Cavuoti},
  {Chabaud}, {Chambers}, {Charles}, {Charlot}, {Chartab}, {Chary}, {Chaumeil},
  {Cho}, {Chon}, {Ciancetta}, {Ciliegi}, {Cimatti}, {Cimino}, {Cioni},
  {Claydon}, {Cleland}, {Cl{\'e}ment}, {Clements}, {Clerc}, {Clesse}, {Codis},
  {Cogato}, {Colbert}, {Cole}, {Coles}, {Collett}, {Collins}, {Colodro-Conde},
  {Colombo}, {Combes}, {Conforti}, {Congedo}, {Conseil}, {Conselice},
  {Contarini}, {Contini}, {Conversi}, {Cooray}, {Copin}, {Corasaniti},
  {Corcho-Caballero}, {Corcione}, {Cordes}, {Corpace}, {Correnti}, {Costanzi},
  {Costille}, {Courbin}, {Courcoult Mifsud}, {Courtois}, {Cousinou}, {Covone},
  {Cowell}, {Cragg}, {Cresci}, {Cristiani}, {Crocce}, {Cropper}, {E Crouzet},
  {Csizi}, {Cuby}, {Cucchetti}, {Cucciati}, {Cuillandre}, {Cunha}, {Cuozzo},
  {Daddi}, {D'Addona}, {Dafonte}, {Dagoneau}, {Dalessandro}, {Dalton},
  {D'Amico}, {Dannerbauer}, {Danto}, {Das}, {Da Silva}, {da Silva},
  {d'Assignies Doumerg}, {Daste}, {Davies}, {Davini}, {Dayal}, {de Boer},
  {Decarli}, {De Caro}, {Degaudenzi}, {Degni}, {de Jong}, {de la Bella}, {de la
  Torre}, {Delhaise}, {Delley}, {Delucchi}, {De Lucia}, {Denniston}, {De
  Paolis}, {De Petris}, {Derosa}, {Desai}, {Desjacques}, {Despali}, {Desprez},
  {De Vicente-Albendea}, {Deville}, {Dias}, {D{\'\i}az-S{\'a}nchez}, {Diaz},
  {Di Domizio}, {Diego}, {Di Ferdinando}, {Di Giorgio}, {Dimauro}, {Dinis},
  {Dolag}, {Dolding}, {Dole}, {Dom{\'\i}nguez S{\'a}nchez}, {Dor{\'e}},
  {Dournac}, {Douspis}, {Dreihahn}, {Droge}, {Dryer}, {Dubath}, {Duc},
  {Ducret}, {Duffy}, {Dufresne}, {Duncan}, {Dupac}, {Duret}, {Durrer},
  {Durret}, {Dusini}, {Ealet}, {Eggemeier}, {Eisenhardt}, {Elbaz}, {Elkhashab},
  {Ellien}, {Endicott}, {Enia}, {Erben}, {Escartin Vigo}, {Escoffier},
  {Escudero Sanz}, {Essert}, {Ettori}, {Ezziati}, {Fabbian}, {Fabricius},
  {Fang}, {Farina}, {Farina}, {Farinelli}, {Farrens}, {Faustini}, {Feltre},
  {Ferguson}, {Ferrando}, {Ferrari}, {Ferr{\'e}-Mateu}, {Ferreira}, {Ferreras},
  {Ferrero}, {Ferriol}, {Ferruit}, {Filleul}, {Finelli}, {Finkelstein},
  {Finoguenov}, {Fiorini}, {Flentge}, {Focardi}, {Fonseca}, {Fontana},
  {Fontanot}, {Fornari}, {Fosalba}, {Fossati}, {Fotopoulou}, {Fouchez},
  {Fourmanoit}, {Frailis}, {Fraix-Burnet}, {Franceschi}, {Franco}, {Franzetti},
  {Freihoefer}, {Frenk}, {Frittoli}, {Frugier}, {Frusciante}, {Fumagalli},
  {Fumagalli}, {Fumana}, {Fu}, {Gabarra}, {Galeotta}, {Galluccio}, {Ganga},
  {Gao}, {Garc{\'\i}a-Bellido}, {Garcia}, {Gardner}, {Garilli},
  {Gaspar-Venancio}, {Gasparetto}, {Gautard}, {Gavazzi}, {Gaztanaga},
  {Genolet}, {Genova Santos}, {Gentile}, {George}, {Gerbino}, {Ghaffari},
  {Giacomini}, {Gianotti}, {Gibb}, {Gillard}, {Gillis}, {Ginolfi}, {Giocoli},
  {Girardi}, {Giri}, {Goh}, {G{\'o}mez-Alvarez}, {Gonzalez-Perez}, {Gonzalez},
  {Gonzalez}, {Gonzalez}, {Gouyou Beauchamps}, {Gozaliasl}, {Gracia-Carpio},
  {Grandis}, {Granett}, {Granvik}, {Grazian}, {Gregorio}, {Grenet}, {Grillo},
  {Grupp}, {Gruppioni}, {Gruppuso}, {Guerbuez}, {Guerrini}, {Guidi},
  {Guillard}, {Gutierrez}, {Guttridge}, {Guzzo}, {Gwyn}, {Haapala}, {Haase},
  {Haddow}, {Hailey}, {Hall}, {Hall}, {Hamaus}, {Haridasu},
  {Harnois-D{\'e}raps}, {Harper}, {Hartley}, {Hasinger}, {Hassani}, {Hatch},
  {Haugan}, {H{\"a}u{\ss}ler}, {Heavens}, {Heisenberg}, {Helmi}, {Helou},
  {Hemmati}, {Henares}, {Herent}, {Hern{\'a}ndez-Monteagudo}, {Heuberger},
  {Hewett}, {Heydenreich}, {Hildebrandt}, {Hirschmann}, {Hjorth}, {Hoar},
  {Hoekstra}, {Holland}, {Holliman}, {Holmes}, {Hook}, {Horeau}, {Hormuth},
  {Hornstrup}, {Hosseini}, {Hu}, {Hudelot}, {Hudson}, {Huertas-Company},
  {Huff}, {Hughes}, {Humphrey}, {Hunt}, {Huynh}, {Ibata}, {Ichikawa},
  {Iglesias-Groth}, {Ilbert}, {Ili{\'c}}, {Ingoglia}, {Iodice}, {Israel},
  {Israelsson}, {Izzo}, {Jablonka}, {Jackson}, {Jacobson}, {Jafariyazani},
  {Jahnke}, {Jain}, {Jansen}, {Jarvis}, {Jasche}, {Jauzac}, {Jeffrey},
  {Jhabvala}, {Jimenez-Teja}, {Jimenez Mu{\~n}oz}, {Joachimi}, {Johansson},
  {Joudaki}, {Jullo}, {Kajava}, {Kang}, {Kannawadi}, {Kansal}, {Karagiannis},
  {K{\"a}rcher}, {Kashlinsky}, {Kazandjian}, {Keck}, {Keih{\"a}nen}, {Kerins},
  {Kermiche}, {Khalil}, {Kiessling}, {Kiiveri}, {Kilbinger}, {Kim}, {King},
  {Kirkpatrick}, {Kitching}, {Kluge}, {Knabenhans}, {Knapen}, {Knebe}, {Kneib},
  {Kohley}, {Koopmans}, {Koskinen}, {Koulouridis}, {Kou}, {Kov{\'a}cs},
  {Kova{\v{c}}i{\'c}}, {Kowalczyk}, {Koyama}, {Kraljic}, {Krause}, {Kruk},
  {Kubik}, {Kuchner}, {Kuijken}, {K{\"u}mmel}, {Kunz}, {Kurki-Suonio},
  {Lacasa}, {Lacey}, {La Franca}, {Lagarde}, {Lahav}, {Laigle}, {La Marca}, {La
  Marle}, {Lamine}, {Lam}, {Lan{\c{c}}on}, {Landt}, {Langer}, {Lapi},
  {Larcheveque}, {Larsen}, {Lattanzi}, {Laudisio}, {Laugier}, {Laureijs},
  {Laurent}, {Lavaux}, {Lawrenson}, {Lazanu}, {Lazeyras}, {Le Boulc'h}, {Le
  Brun}, {Le Brun}, {Leclercq}, {Lee}, {Le Graet}, {Legrand}, {Leirvik}, {Le
  Jeune}, {Lembo}, {Le Mignant}, {Lepinzan}, {Lepori}, {Le Reun}, {Leroy},
  {Lesci}, {Lesgourgues}, {Leuzzi}, {Levi}, {Liaudat}, {Libet}, {Liebing},
  {Ligori}, {Lilje}, {Lin}, {Linde}, {Linder}, {Lindholm}, {Linke}, {Li},
  {Liu}, {Lloro}, {Lobo}, {Lodieu}, {Lombardi}, {Lombriser}, {Lonare}, {Longo},
  {L{\'o}pez-Caniego}, {Lopez Lopez}, {Alvarez}, {Loureiro}, {Loveday},
  {Lusso}, {Macias-Perez}, {Maciaszek}, {Maggio}, {Magliocchetti}, {Magnard},
  {Magnier}, {Magro}, {Mahler}, {Mainetti}, {Maino}, {Maiorano}, {Maiorano},
  {Malavasi}, {Mamon}, {Mancini}, {Mandelbaum}, {Manera},
  {Manj{\'o}n-Garc{\'\i}a}, {Mannucci}, {Mansutti}, {Manteiga Outeiro},
  {Maoli}, {Maraston}, {Marcin}, {Marcos-Arenal}, {Margalef-Bentabol},
  {Marggraf}, {Marinucci}, {Marinucci}, {Markovic}, {Marleau}, {Marpaud},
  {Martignac}, {Mart{\'\i}n-Fleitas}, {Martin-Moruno}, {Martin}, {Martinelli},
  {Martinet}, {Martin}, {Martins}, {Marulli}, {Massari}, {Massey}, {Masters},
  {Matarrese}, {Matsuoka}, {Matthew}, {Maughan}, {Mauri}, {Maurin},
  {Maurogordato}, {McCarthy}, {McConnachie}, {McCracken}, {McDonald}, {McEwen},
  {McPartland}, {Medinaceli}, {Mehta}, {Mei}, {Melchior}, {Melin},
  {M{\'e}nard}, {Mendes}, {Mendez-Abreu}, {Meneghetti}, {Mercurio}, {Merlin},
  {Metcalf}, {Meylan}, {Migliaccio}, {Mignoli}, {Miller}, {Miluzio},
  {Milvang-Jensen}, {Mimoso}, {Miquel}, {Miyatake}, {Mobasher}, {Mohr},
  {Monaco}, {Mongui{\'o}}, {Montoro}, {Mora}, {Moradinezhad Dizgah}, {Moresco},
  {Moretti}, {Morgante}, {Morisset}, {Moriya}, {Morris}, {Mortlock},
  {Moscardini}, {Mota}, {Mottet}, {Moustakas}, {Moutard}, {M{\"u}ller},
  {Munari}, {Murphree}, {Murray}, {Murray}, {Musi}, {Nadathur}, {Nagam},
  {Nagao}, {Naidoo}, {Nakajima}, {Nally}, {Natoli}, {Navarro-Alsina}, {Navarro
  Girones}, {Neissner}, {Nersesian}, {Nesseris}, {Nguyen-Kim}, {Nicastro},
  {Nichol}, {Nielbock}, {Niemi}, {Nieto}, {Nilsson}, {Noller}, {Norberg},
  {Nouri-Zonoz}, {Ntelis}, {Nucita}, {Nugent}, {Nunes}, {Nutma}, {Ocampo},
  {Odier}, {Oesch}, {Oguri}, {Magalhaes Oliveira}, {Onoue}, {Oosterbroek},
  {Oppizzi}, {Ordenovic}, {Osato}, {Pacaud}, {Pace}, {Padilla}, {Paech},
  {Pagano}, {Page}, {Palazzi}, {Paltani}, {Pamuk}, {Pandolfi}, {Paoletti},
  {Paolillo}, {Papaderos}, {Pardede}, {Parimbelli}, {Parmar}, {Partmann},
  {Pasian}, {Passalacqua}, {Paterson}, {Patrizii}, {Pattison},
  {Paulino-Afonso}, {Paviot}, {Peacock}, {Pearce}, {Pedersen}, {Peel},
  {Peletier}, {Pellejero Ibanez}, {Pello}, {Penny}, {Percival},
  {Perez-Garrido}, {Perotto}, {Pettorino}, {Pezzotta}, {Pezzuto}, {Philippon},
  {Pierre}, {Piersanti}, {Pietroni}, {Piga}, {Pilo}, {Pires}, {Pisani},
  {Pizzella}, {Pizzuti}, {Plana}, {Polenta}, {Pollack}, {Poncet},
  {P{\"o}ntinen}, {Pool}, {Popa}, {Popa}, {Popp}, {Porciani}, {Porth},
  {Potter}, {Poulain}, {Pourtsidou}, {Pozzetti}, {Prandoni}, {Pratt},
  {Prezelus}, {Prieto}, {Pugno}, {Quai}, {Quilley}, {Racca}, {Raccanelli},
  {R{\'a}cz}, {Radinovi{\'c}}, {Radovich}, {Ragagnin}, {Ragnit}, {Raison},
  {Ramos-Chernenko}, {Ranc}, {Rasera}, {Raylet}, {Rebolo}, {Refregier},
  {Reimberg}, {Reiprich}, {Renk}, {Renzi}, {Retre}, {Revaz}, {Reyl{\'e}},
  {Reynolds}, {Rhodes}, {Ricci}, {Ricci}, {Riccio}, {Ricken}, {Rissanen},
  {Risso}, {Rix}, {Robin}, {Rocca-Volmerange}, {Rocci}, {Rodenhuis},
  {Rodighiero}, {Rodriguez Monroy}, {Rollins}, {Romanello}, {Roman}, {Romelli},
  {Romero-Gomez}, {Roncarelli}, {Rosati}, {Rosset}, {Rossetti}, {Roster},
  {Rottgering}, {Rozas-Fern{\'a}ndez}, {Ruane}, {Rubino-Martin}, {Rudolph},
  {Ruppin}, {Rusholme}, {Sacquegna}, {S{\'a}ez-Casares}, {Saga}, {Saglia},
  {Sahl{\'e}n}, {Saifollahi}, {Sakr}, {Salvalaggio}, {Salvaterra}, {Salvati},
  {Salvato}, {Salvignol}, {S{\'a}nchez}, {Sanchez}, {Sanders}, {Sapone},
  {Saponara}, {Sarpa}, {Sarron}, {Sartori}, {Sartoris}, {Sassolas}, {Sauniere},
  {Sauvage}, {Sawicki}, {Scaramella}, {Scarlata}, {Scharr{\'e}}, {Schaye},
  {Schewtschenko}, {Schindler}, {Schinnerer}, {Schirmer}, {Schmidt}, {Schmidt},
  {Schmidt}, {Schneider}, {Schneider}, {Schneider}, {Sch{\"o}neberg},
  {Schrabback}, {Schultheis}, {Schulz}, {Schuster}, {Schwartz}, {Sciotti},
  {Scodeggio}, {Scognamiglio}, {Scott}, {Scottez}, {Secroun}, {Sefusatti},
  {Seidel}, {Seiffert}, {Sellentin}, {Selwood}, {Semboloni}, {Sereno},
  {Serjeant}, {Serrano}, {Setnikar}, {Shankar}, {Sharples}, {Short},
  {Shulevski}, {Shuntov}, {Sias}, {Sikkema}, {Silvestri}, {Simon}, {Sirignano},
  {Sirri}, {Skottfelt}, {Slezak}, {Sluse}, {Smith}, {Smith}, {Smith}, {Smit},
  {Soldano}, {Solheim}, {Sorce}, {Sorrenti}, {Soubrie}, {Spinoglio}, {Spurio
  Mancini}, {Stadel}, {Stagnaro}, {Stanco}, {Stanford}, {Starck}, {Stassi},
  {Steinwagner}, {Stern}, {Stone}, {Strada}, {Strafella}, {Stramaccioni},
  {Surace}, {Sureau}, {Suyu}, {Swindells}, {Szafraniec}, {Szapudi}, {Taamoli},
  {Talia}, {Tallada-Cresp{\'\i}}, {Tanidis}, {Tao}, {Tarr{\'\i}o},
  {Tavagnacco}, {Taylor}, {Taylor}, {Taylor}, {Teixeira}, {Tenti}, {Teodoro
  Idiago}, {Teplitz}, {Tereno}, {Tessore}, {Testa}, {Testera}, {Tewes},
  {Teyssier}, {Theret}, {Thizy}, {Thomas}, {Toba}, {Toft}, {Toledo-Moreo},
  {Tolstoy}, {Tommasi}, {Torbaniuk}, {Torradeflot}, {Tortora}, {Tosi}, {Tosti},
  {Trifoglio}, {Troja}, {Trombetti}, {Tronconi}, {Tsedrik}, {Tsyganov},
  {Tucci}, {Tutusaus}, {Uhlemann}, {Ulivi}, {Urbano}, {Vacher}, {Vaillon},
  {Valageas}, {Valdes}, {Valentijn}, {Valenziano}, {Valieri}, {Valiviita}, {Van
  den Broeck}, {Vassallo}, {Vavrek}, {Vega-Ferrero}, {Venemans}, {Venhola},
  {Ventura}, {Verdoes Kleijn}, {Vergani}, {Verma}, {Vernizzi}, {Veropalumbo},
  {Verza}, {Vescovi}, {Vibert}, {Viel}, {Vielzeuf}, {Viglione}, {Viitanen},
  {Villaescusa-Navarro}, {Vinciguerra}, {Visticot}, {Voggel}, {von
  Wietersheim-Kramsta}, {Vriend}, {Wachter}, {Walmsley}, {Walth}, {Walton},
  {Walton}, {Wander}, {Wang}, {Wang}, {Weaver}, {Weller}, {Wetzstein},
  {Whalen}, {Whittam}, {Widmer}, {Wiesmann}, {Wilde}, {Williams}, {Winther},
  {Wittje}, {Wong}, {Wright}, {Yankelevich}, {Yeung}, {Yoon}, {Youles}, {Yung},
  {Zacchei}, {Zalesky}, {Zamorani}, {Zamorano Vitorelli}, {Zanoni Marc},
  {Zennaro}, {Zerbi}, {Zinchenko}, {Zoubian}, {Zucca}, \&
  {Zumalacarregui}}]{euclid}
{Euclid Collaboration}, {Mellier}, Y., {Abdurro'uf}, {et~al.} 2024, arXiv
  e-prints, arXiv:2405.13491, \dodoi{10.48550/arXiv.2405.13491}

\bibitem[{{Finkelstein} {et~al.}(2023){Finkelstein}, {Bagley}, {Ferguson},
  {Wilkins}, {Kartaltepe}, {Papovich}, {Yung}, {Haro}, {Behroozi}, {Dickinson},
  {Kocevski}, {Koekemoer}, {Larson}, {Le Bail}, {Morales},
  {P{\'e}rez-Gonz{\'a}lez}, {Burgarella}, {Dav{\'e}}, {Hirschmann},
  {Somerville}, {Wuyts}, {Bromm}, {Casey}, {Fontana}, {Fujimoto}, {Gardner},
  {Giavalisco}, {Grazian}, {Grogin}, {Hathi}, {Hutchison}, {Jha}, {Jogee},
  {Kewley}, {Kirkpatrick}, {Long}, {Lotz}, {Pentericci}, {Pierel}, {Pirzkal},
  {Ravindranath}, {Ryan}, {Trump}, {Yang}, {Bhatawdekar}, {Bisigello}, {Buat},
  {Calabr{\`o}}, {Castellano}, {Cleri}, {Cooper}, {Croton}, {Daddi}, {Dekel},
  {Elbaz}, {Franco}, {Gawiser}, {Holwerda}, {Huertas-Company}, {Jaskot},
  {Leung}, {Lucas}, {Mobasher}, {Pandya}, {Tacchella}, {Weiner}, \&
  {Zavala}}]{CEERS}
{Finkelstein}, S.~L., {Bagley}, M.~B., {Ferguson}, H.~C., {et~al.} 2023, \apjl,
  946, L13, \dodoi{10.3847/2041-8213/acade4}

\bibitem[{{Gardner} {et~al.}(2006){Gardner}, {Mather}, {Clampin}, {Doyon},
  {Greenhouse}, {Hammel}, {Hutchings}, {Jakobsen}, {Lilly}, {Long}, {Lunine},
  {McCaughrean}, {Mountain}, {Nella}, {Rieke}, {Rieke}, {Rix}, {Smith},
  {Sonneborn}, {Stiavelli}, {Stockman}, {Windhorst}, \&
  {Wright}}]{Gardner2006SSRv}
{Gardner}, J.~P., {Mather}, J.~C., {Clampin}, M., {et~al.} 2006, \ssr, 123,
  485, \dodoi{10.1007/s11214-006-8315-7}

\bibitem[{{Guo} {et~al.}(2013){Guo}, {Ferguson}, {Giavalisco}, {Barro},
  {Willner}, {Ashby}, {Dahlen}, {Donley}, {Faber}, {Fontana}, {Galametz},
  {Grazian}, {Huang}, {Kocevski}, {Koekemoer}, {Koo}, {McGrath}, {Peth},
  {Salvato}, {Wuyts}, {Castellano}, {Cooray}, {Dickinson}, {Dunlop}, {Fazio},
  {Gardner}, {Gawiser}, {Grogin}, {Hathi}, {Hsu}, {Lee}, {Lucas}, {Mobasher},
  {Nandra}, {Newman}, \& {van der Wel}}]{2013ApJS..207...24G}
{Guo}, Y., {Ferguson}, H.~C., {Giavalisco}, M., {et~al.} 2013, \apjs, 207, 24,
  \dodoi{10.1088/0067-0049/207/2/24}

\bibitem[{Harris {et~al.}(2020)Harris, Millman, van~der Walt, Gommers,
  Virtanen, Cournapeau, Wieser, Taylor, Berg, Smith, Kern, Picus, Hoyer, van
  Kerkwijk, Brett, Haldane, del R{\'{i}}o, Wiebe, Peterson,
  G{\'{e}}rard-Marchant, Sheppard, Reddy, Weckesser, Abbasi, Gohlke, \&
  Oliphant}]{Harris2020}
Harris, C.~R., Millman, K.~J., van~der Walt, S.~J., {et~al.} 2020, Nature, 585,
  357, \dodoi{10.1038/s41586-020-2649-2}

\bibitem[{Hausen \& Robertson(2022)}]{hausen2022a}
Hausen, R., \& Robertson, B. 2022, Astronomy and Computing, 39, 100586,
  \dodoi{https://doi.org/10.1016/j.ascom.2022.100586}

\bibitem[{{Horne}(1986)}]{Horne86}
{Horne}, K. 1986, \pasp, 98, 609, \dodoi{10.1086/131801}

\bibitem[{Hunter(2007)}]{Hunter2007}
Hunter, J.~D. 2007, Computing in Science \& Engineering, 9, 90,
  \dodoi{10.1109/MCSE.2007.55}

\bibitem[{{Kashino} {et~al.}(2023){Kashino}, {Lilly}, {Matthee}, {Eilers},
  {Mackenzie}, {Bordoloi}, \& {Simcoe}}]{EIGER}
{Kashino}, D., {Lilly}, S.~J., {Matthee}, J., {et~al.} 2023, \apj, 950, 66,
  \dodoi{10.3847/1538-4357/acc588}

\bibitem[{{Kron}(1980)}]{Kron1980}
{Kron}, R.~G. 1980, \apjs, 43, 305, \dodoi{10.1086/190669}

\bibitem[{{Labb{\'e}} {et~al.}(2005){Labb{\'e}}, {Huang}, {Franx}, {Rudnick},
  {Barmby}, {Daddi}, {van Dokkum}, {Fazio}, {F{\"o}rster Schreiber},
  {Moorwood}, {Rix}, {R{\"o}ttgering}, {Trujillo}, \& {van der
  Werf}}]{2005ApJ...624L..81L}
{Labb{\'e}}, I., {Huang}, J., {Franx}, M., {et~al.} 2005, \apjl, 624, L81,
  \dodoi{10.1086/430700}

\bibitem[{{Maiolino} {et~al.}(2024){Maiolino}, {Scholtz}, {Witstok},
  {Carniani}, {D'Eugenio}, {de Graaff}, {{\"U}bler}, {Tacchella},
  {Curtis-Lake}, {Arribas}, {Bunker}, {Charlot}, {Chevallard}, {Curti},
  {Looser}, {Maseda}, {Rawle}, {Rodr{\'\i}guez del Pino}, {Willott}, {Egami},
  {Eisenstein}, {Hainline}, {Robertson}, {Williams}, {Willmer}, {Baker},
  {Boyett}, {DeCoursey}, {Fabian}, {Helton}, {Ji}, {Jones}, {Kumari},
  {Laporte}, {Nelson}, {Perna}, {Sandles}, {Shivaei}, \&
  {Sun}}]{2024Natur.627...59M}
{Maiolino}, R., {Scholtz}, J., {Witstok}, J., {et~al.} 2024, \nat, 627, 59,
  \dodoi{10.1038/s41586-024-07052-5}

\bibitem[{{Merlin} {et~al.}(2024){Merlin}, {Santini}, {Paris}, {Castellano},
  {Fontana}, {Treu}, {Finkelstein}, {Dunlop}, {Arrabal Haro}, {Bagley},
  {Boyett}, {Calabr{\`o}}, {Correnti}, {Davis}, {Dickinson}, {Donnan},
  {Ferguson}, {Fortuni}, {Giavalisco}, {Glazebrook}, {Grazian}, {Grogin},
  {Hathi}, {Hirschmann}, {Kartaltepe}, {Kewley}, {Kirkpatrick}, {Kocevski},
  {Koekemoer}, {Leung}, {Lotz}, {Lucas}, {Magee}, {Marchesini}, {Mascia},
  {McLeod}, {McLure}, {Nanayakkara}, {Napolitano}, {Nonino}, {Papovich},
  {Pentericci}, {P{\'e}rez-Gonz{\'a}lez}, {Pirzkal}, {Ravindranath},
  {Roberts-Borsani}, {Somerville}, {Trenti}, {Trump}, {Vulcani}, {Wang},
  {Watson}, {Wilkins}, {Yang}, \& {Yung}}]{2024arXiv240900169M}
{Merlin}, E., {Santini}, P., {Paris}, D., {et~al.} 2024, arXiv e-prints,
  arXiv:2409.00169, \dodoi{10.48550/arXiv.2409.00169}

\bibitem[{{Meyer} {et~al.}(2024){Meyer}, {Oesch}, {Giovinazzo}, {Weibel},
  {Brammer}, {Matthee}, {Naidu}, {Bouwens}, {Chisholm}, {Covelo-Paz},
  {Fudamoto}, {Maseda}, {Nelson}, {Shivaei}, {Xiao}, {Herard-Demanche},
  {Illingworth}, {Kerutt}, {Kramarenko}, {Labbe}, {Leonova}, {Magee},
  {Matharu}, {Lyon}, {Reddy}, {Schaerer}, {Shapley}, {Stefanon}, {Wozniak}, \&
  {Wuyts}}]{Meyer24}
{Meyer}, R.~A., {Oesch}, P.~A., {Giovinazzo}, E., {et~al.} 2024, \mnras,
  \dodoi{10.1093/mnras/stae2353}

\bibitem[{{Oesch} {et~al.}(2023){Oesch}, {Brammer}, {Naidu}, {Bouwens},
  {Chisholm}, {Illingworth}, {Matthee}, {Nelson}, {Qin}, {Reddy}, {Shapley},
  {Shivaei}, {van Dokkum}, {Weibel}, {Whitaker}, {Wuyts}, {Covelo-Paz},
  {Endsley}, {Fudamoto}, {Giovinazzo}, {Herard-Demanche}, {Kerutt},
  {Kramarenko}, {Labbe}, {Leonova}, {Lin}, {Magee}, {Marchesini}, {Maseda},
  {Mason}, {Matharu}, {Meyer}, {Neufeld}, {Prieto Lyon}, {Schaerer}, {Sharma},
  {Shuntov}, {Smit}, {Stefanon}, {Wyithe}, \& {Xiao}}]{FRESCO}
{Oesch}, P.~A., {Brammer}, G., {Naidu}, R.~P., {et~al.} 2023, \mnras, 525,
  2864, \dodoi{10.1093/mnras/stad2411}

\bibitem[{{Rieke} {et~al.}(2023{\natexlab{a}}){Rieke}, {Kelly}, {Misselt},
  {Stansberry}, {Boyer}, {Beatty}, {Egami}, {Florian}, {Greene}, {Hainline},
  {Leisenring}, {Roellig}, {Schlawin}, {Sun}, {Tinnin}, {Williams}, {Willmer},
  {Wilson}, {Clark}, {Rohrbach}, {Brooks}, {Canipe}, {Correnti}, {DiFelice},
  {Gennaro}, {Girard}, {Hartig}, {Hilbert}, {Koekemoer}, {Nikolov}, {Pirzkal},
  {Rest}, {Robberto}, {Sunnquist}, {Telfer}, {Wu}, {Ferry}, {Lewis}, {Baum},
  {Beichman}, {Doyon}, {Dressler}, {Eisenstein}, {Ferrarese}, {Hodapp},
  {Horner}, {Jaffe}, {Johnstone}, {Krist}, {Martin}, {McCarthy}, {Meyer},
  {Rieke}, {Trauger}, \& {Young}}]{Rieke23}
{Rieke}, M.~J., {Kelly}, D.~M., {Misselt}, K., {et~al.} 2023{\natexlab{a}},
  \pasp, 135, 028001, \dodoi{10.1088/1538-3873/acac53}

\bibitem[{{Rieke} {et~al.}(2023{\natexlab{b}}){Rieke}, {Robertson},
  {Tacchella}, {Hainline}, {Johnson}, {Hausen}, {Ji}, {Willmer}, {Eisenstein},
  {Pusk{\'a}s}, {Alberts}, {Arribas}, {Baker}, {Baum}, {Bhatawdekar},
  {Bonaventura}, {Boyett}, {Bunker}, {Cameron}, {Carniani}, {Charlot},
  {Chevallard}, {Chen}, {Curti}, {Curtis-Lake}, {Danhaive}, {DeCoursey},
  {Dressler}, {Egami}, {Endsley}, {Helton}, {Hviding}, {Kumari}, {Looser},
  {Lyu}, {Maiolino}, {Maseda}, {Nelson}, {Rieke}, {Rix}, {Sandles}, {Saxena},
  {Sharpe}, {Shivaei}, {Skarbinski}, {Smit}, {Stark}, {Stone}, {Suess}, {Sun},
  {Topping}, {{\"U}bler}, {Villanueva}, {Wallace}, {Williams}, {Willott},
  {Whitler}, {Witstok}, \& {Woodrum}}]{2023ApJS..269...16R}
{Rieke}, M.~J., {Robertson}, B., {Tacchella}, S., {et~al.} 2023{\natexlab{b}},
  \apjs, 269, 16, \dodoi{10.3847/1538-4365/acf44d}

\bibitem[{{Rigby} {et~al.}(2023){Rigby}, {Perrin}, {McElwain}, {Kimble},
  {Friedman}, {Lallo}, {Doyon}, {Feinberg}, {Ferruit}, {Glasse}, \&
  et~al.}]{Rigby23}
{Rigby}, J., {Perrin}, M., {McElwain}, M., {et~al.} 2023, \pasp, 135, 048001,
  \dodoi{10.1088/1538-3873/acb293}

\bibitem[{{Schouws} {et~al.}(2024){Schouws}, {Bouwens}, {Ormerod}, {Smit},
  {Algera}, {Sommovigo}, {Hodge}, {Ferrara}, {Oesch}, {Rowland}, {van Leeuwen},
  {Stefanon}, {Herard-Demanche}, {Fudamoto}, {R{\"o}ttgering}, \& {van der
  Werf}}]{2024arXiv240920549S}
{Schouws}, S., {Bouwens}, R.~J., {Ormerod}, K., {et~al.} 2024, arXiv e-prints,
  arXiv:2409.20549, \dodoi{10.48550/arXiv.2409.20549}

\bibitem[{{Shen} {et~al.}(2024){Shen}, {Zhuang}, {Li}, {Burgasser}, {Fan},
  {Greene}, {Narayan}, {Shapley}, {Sun}, {Wang}, \& {Yang}}]{nexus}
{Shen}, Y., {Zhuang}, M.-Y., {Li}, J., {et~al.} 2024, arXiv e-prints,
  arXiv:2408.12713, \dodoi{10.48550/arXiv.2408.12713}

\bibitem[{{Skelton} {et~al.}(2014){Skelton}, {Whitaker}, {Momcheva}, {Brammer},
  {van Dokkum}, {Labb{\'e}}, {Franx}, {van der Wel}, {Bezanson}, {Da Cunha},
  {Fumagalli}, {F{\"o}rster Schreiber}, {Kriek}, {Leja}, {Lundgren}, {Magee},
  {Marchesini}, {Maseda}, {Nelson}, {Oesch}, {Pacifici}, {Patel}, {Price},
  {Rix}, {Tal}, {Wake}, \& {Wuyts}}]{2014ApJS..214...24S}
{Skelton}, R.~E., {Whitaker}, K.~E., {Momcheva}, I.~G., {et~al.} 2014, \apjs,
  214, 24, \dodoi{10.1088/0067-0049/214/2/24}

\bibitem[{{Sun} {et~al.}(2022){Sun}, {Egami}, {Pirzkal}, {Rieke}, {Boyer},
  {Correnti}, {Gennaro}, {Girard}, {Greene}, {Kelly}, {Koekemoer},
  {Leisenring}, {Misselt}, {Nikolov}, {Roellig}, {Stansberry}, {Williams},
  {Willmer}, \& {Members of the JWST/NIRCam Commissioning Team}}]{Sun22}
{Sun}, F., {Egami}, E., {Pirzkal}, N., {et~al.} 2022, \apjl, 936, L8,
  \dodoi{10.3847/2041-8213/ac8938}

\bibitem[{{Sun} {et~al.}(2023){Sun}, {Egami}, {Pirzkal}, {Rieke}, {Baum},
  {Boyer}, {Boyett}, {Bunker}, {Cameron}, {Curti}, {Eisenstein}, {Gennaro},
  {Greene}, {Jaffe}, {Kelly}, {Koekemoer}, {Kumari}, {Maiolino}, {Maseda},
  {Perna}, {Rest}, {Robertson}, {Schlawin}, {Smit}, {Stansberry}, {Sunnquist},
  {Tacchella}, {Williams}, \& {Willmer}}]{Sun23}
---. 2023, \apj, 953, 53, \dodoi{10.3847/1538-4357/acd53c}

\bibitem[{{Taylor} {et~al.}(2023){Taylor}, {Barger}, {Cowie}, {Hasinger}, {Hu},
  \& {Songaila}}]{2023ApJS..266...24T}
{Taylor}, A.~J., {Barger}, A.~J., {Cowie}, L.~L., {et~al.} 2023, \apjs, 266,
  24, \dodoi{10.3847/1538-4365/accd70}

\bibitem[{{Treu} {et~al.}(2022){Treu}, {Roberts-Borsani}, {Bradac}, {Brammer},
  {Fontana}, {Henry}, {Mason}, {Morishita}, {Pentericci}, {Wang}, {Acebron},
  {Bagley}, {Bergamini}, {Belfiori}, {Bonchi}, {Boyett}, {Boutsia},
  {Calabr{\'o}}, {Caminha}, {Castellano}, {Dressler}, {Glazebrook}, {Grillo},
  {Jacobs}, {Jones}, {Kelly}, {Leethochawalit}, {Malkan}, {Marchesini},
  {Mascia}, {Mercurio}, {Merlin}, {Nanayakkara}, {Nonino}, {Paris},
  {Poggianti}, {Rosati}, {Santini}, {Scarlata}, {Shipley}, {Strait}, {Trenti},
  {Tubthong}, {Vanzella}, {Vulcani}, \& {Yang}}]{GLASS}
{Treu}, T., {Roberts-Borsani}, G., {Bradac}, M., {et~al.} 2022, \apj, 935, 110,
  \dodoi{10.3847/1538-4357/ac8158}

\bibitem[{{Virtanen} {et~al.}(2020){Virtanen}, {Gommers}, {Oliphant},
  {Haberland}, {Reddy}, {Cournapeau}, {Burovski}, {Peterson}, {Weckesser},
  {Bright}, {van der Walt}, {Brett}, {Wilson}, {Millman}, {Mayorov}, {Nelson},
  {Jones}, {Kern}, {Larson}, {Carey}, {Polat}, {Feng}, {Moore}, {VanderPlas},
  {Laxalde}, {Perktold}, {Cimrman}, {Henriksen}, {Quintero}, {Harris},
  {Archibald}, {Ribeiro}, {Pedregosa}, {van Mulbregt}, \& {SciPy 1. 0
  Contributors}}]{scipy}
{Virtanen}, P., {Gommers}, R., {Oliphant}, T.~E., {et~al.} 2020, Nature
  Methods, 17, 261, \dodoi{10.1038/s41592-019-0686-2}

\bibitem[{{Wang} {et~al.}(2023){Wang}, {Yang}, {Hennawi}, {Fan}, {Sun},
  {Champagne}, {Costa}, {Habouzit}, {Endsley}, {Li}, {Lin}, {Meyer},
  {Schindler}, {Wu}, {Ba{\~n}ados}, {Barth}, {Bhowmick}, {Bieri}, {Blecha},
  {Bosman}, {Cai}, {Colina}, {Connor}, {Davies}, {Decarli}, {De Rosa}, {Drake},
  {Egami}, {Eilers}, {Evans}, {Farina}, {Haiman}, {Jiang}, {Jin}, {Jun},
  {Kakiichi}, {Khusanova}, {Kulkarni}, {Li}, {Liu}, {Loiacono}, {Lupi},
  {Mazzucchelli}, {Onoue}, {Pudoka}, {Rojas-Ruiz}, {Shen}, {Strauss}, {Tee},
  {Trakhtenbrot}, {Trebitsch}, {Venemans}, {Volonteri}, {Walter}, {Xie}, {Yue},
  {Zhang}, {Zhang}, \& {Zou}}]{Wang23}
{Wang}, F., {Yang}, J., {Hennawi}, J.~F., {et~al.} 2023, \apjl, 951, L4,
  \dodoi{10.3847/2041-8213/accd6f}

\bibitem[{{Whitaker} {et~al.}(2011){Whitaker}, {Labb{\'e}}, {van Dokkum},
  {Brammer}, {Kriek}, {Marchesini}, {Quadri}, {Franx}, {Muzzin}, {Williams},
  {Bezanson}, {Illingworth}, {Lee}, {Lundgren}, {Nelson}, {Rudnick}, {Tal}, \&
  {Wake}}]{2011ApJ...735...86W}
{Whitaker}, K.~E., {Labb{\'e}}, I., {van Dokkum}, P.~G., {et~al.} 2011, \apj,
  735, 86, \dodoi{10.1088/0004-637X/735/2/86}

\bibitem[{{Windhorst} {et~al.}(2023){Windhorst}, {Cohen}, {Jansen}, {Summers},
  {Tompkins}, {Conselice}, {Driver}, {Yan}, {Coe}, {Frye}, {Grogin},
  {Koekemoer}, {Marshall}, {O'Brien}, {Pirzkal}, {Robotham}, {Ryan}, {Willmer},
  {Carleton}, {Diego}, {Keel}, {Porto}, {Redshaw}, {Scheller}, {Wilkins},
  {Willner}, {Zitrin}, {Adams}, {Austin}, {Arendt}, {Beacom}, {Bhatawdekar},
  {Bradley}, {Broadhurst}, {Cheng}, {Civano}, {Dai}, {Dole}, {D'Silva},
  {Duncan}, {Fazio}, {Ferrami}, {Ferreira}, {Finkelstein}, {Furtak}, {Gim},
  {Griffiths}, {Hammel}, {Harrington}, {Hathi}, {Holwerda}, {Honor}, {Huang},
  {Hyun}, {Im}, {Joshi}, {Kamieneski}, {Kelly}, {Larson}, {Li}, {Lim}, {Ma},
  {Maksym}, {Manzoni}, {Meena}, {Milam}, {Nonino}, {Pascale}, {Petric},
  {Pierel}, {del Carmen Polletta}, {R{\"o}ttgering}, {Rutkowski}, {Smail},
  {Straughn}, {Strolger}, {Swirbul}, {Trussler}, {Wang}, {Welch}, {B. Wyithe},
  {Yun}, {Zackrisson}, {Zhang}, \& {Zhao}}]{PEARLS}
{Windhorst}, R.~A., {Cohen}, S.~H., {Jansen}, R.~A., {et~al.} 2023, \aj, 165,
  13, \dodoi{10.3847/1538-3881/aca163}

\bibitem[{{Wright} {et~al.}(2023){Wright}, {Rieke}, {Glasse}, {Ressler},
  {Garc{\'\i}a Mar{\'\i}n}, {Aguilar}, {Alberts}, {{\'A}lvarez-M{\'a}rquez},
  {Argyriou}, {Banks}, {Baudoz}, {Boccaletti}, {Bouchet}, {Bouwman}, {Brandl},
  {Breda}, {Bright}, {Cale}, {Colina}, {Cossou}, {Coulais}, {Cracraft}, {De
  Meester}, {Dicken}, {Engesser}, {Etxaluze}, {Fox}, {Friedman}, {Fu},
  {Gasman}, {G{\'a}sp{\'a}r}, {Gastaud}, {Geers}, {Glauser}, {Gordon},
  {Greene}, {Greve}, {Grundy}, {G{\"u}del}, {Guillard}, {Haderlein},
  {Hashimoto}, {Henning}, {Hines}, {Holler}, {Detre}, {Jahromi}, {James},
  {Jones}, {Justtanont}, {Kavanagh}, {Kendrew}, {Klaassen}, {Krause},
  {Labiano}, {Lagage}, {Lambros}, {Larson}, {Law}, {Lee}, {Libralato}, {Lorenzo
  Alverez}, {Meixner}, {Morrison}, {Mueller}, {Murray}, {Mycroft}, {Myers},
  {Nayak}, {Naylor}, {Nickson}, {Noriega-Crespo}, {{\"O}stlin}, {O'Sullivan},
  {Ottens}, {Patapis}, {Penanen}, {Pietraszkiewicz}, {Ray}, {Regan},
  {Roteliuk}, {Royer}, {Samara-Ratna}, {Samuelson}, {Sargent}, {Scheithauer},
  {Schneider}, {Schreiber}, {Shaughnessy}, {Sheehan}, {Shivaei}, {Sloan},
  {Tamas}, {Teague}, {Temim}, {Tikkanen}, {Tustain}, {van Dishoeck},
  {Vandenbussche}, {Weilert}, {Whitehouse}, \& {Wolff}}]{Wright23}
{Wright}, G.~S., {Rieke}, G.~H., {Glasse}, A., {et~al.} 2023, \pasp, 135,
  048003, \dodoi{10.1088/1538-3873/acbe66}

\bibitem[{{Yang} {et~al.}(2023){Yang}, {Wang}, {Fan}, {Hennawi}, {Barth},
  {Ba{\~n}ados}, {Sun}, {Liu}, {Cai}, {Jiang}, {Li}, {Onoue}, {Schindler},
  {Shen}, {Wu}, {Bhowmick}, {Bieri}, {Blecha}, {Bosman}, {Champagne}, {Colina},
  {Connor}, {Costa}, {Davies}, {Decarli}, {De Rosa}, {Drake}, {Egami},
  {Eilers}, {Evans}, {Farina}, {Habouzit}, {Haiman}, {Jin}, {Jun}, {Kakiichi},
  {Khusanova}, {Kulkarni}, {Loiacono}, {Lupi}, {Mazzucchelli}, {Pan},
  {Rojas-Ruiz}, {Strauss}, {Tee}, {Trakhtenbrot}, {Trebitsch}, {Venemans},
  {Vestergaard}, {Volonteri}, {Walter}, {Xie}, {Yue}, {Zhang}, {Zhang}, \&
  {Zou}}]{Yang23}
{Yang}, J., {Wang}, F., {Fan}, X., {et~al.} 2023, \apjl, 951, L5,
  \dodoi{10.3847/2041-8213/acc9c8}

\bibitem[{{Zhuang} {et~al.}(2024){Zhuang}, {Li}, \&
  {Shen}}]{2024ApJ...962...93Z}
{Zhuang}, M.-Y., {Li}, J., \& {Shen}, Y. 2024, \apj, 962, 93,
  \dodoi{10.3847/1538-4357/ad1517}

\bibitem[{{Zhuang} \& {Shen}(2024)}]{2024ApJ...962..139Z}
{Zhuang}, M.-Y., \& {Shen}, Y. 2024, \apj, 962, 139,
  \dodoi{10.3847/1538-4357/ad1183}

\end{thebibliography}
\bibliographystyle{aasjournal}



\end{document}